\documentclass[journal]{IEEEtran}

\usepackage[
  colorlinks=true,
  linkcolor=blue,
  citecolor=blue,
  urlcolor=blue
]{hyperref}

\usepackage[T1]{fontenc}
\usepackage{times}
\usepackage{amsmath,amssymb,amsfonts}
\usepackage{bm}
\usepackage{graphicx}
\usepackage{xcolor}
\usepackage{cite}
\usepackage{url}
\usepackage{mathtools}
\usepackage{enumitem}
\usepackage{comment}

\newcommand{\Cin}{\mathbf C_{\mathrm{in}}}
\newcommand{\Cout}{\mathbf C_{\mathrm{out}}}
\newcommand{\rhoout}{\boldsymbol\rho_{\mathrm{out}}}
\newcommand{\rhostar}{\boldsymbol\rho_{\star}}
\newcommand{\Hmat}{\mathbf H}
\newcommand{\Sout}{S_{\mathrm{out}}}

\title{Bounds on Shaping Partially Coherent Microwaves with Programmable Scattering Systems}

\author{Philipp~del~Hougne,~\IEEEmembership{Member,~IEEE}
\thanks{
P.~del~Hougne is with Univ Rennes, CNRS, IETR -- UMR 6164, F-35000 Rennes, France. (e-mail: philipp.del-hougne@univ-rennes.fr)
}
\thanks{This work was supported in part by the ANR France 2030 program (project ANR-22-PEFT-0005), the ANR PRCI program (project ANR-22-CE93-0010), the Rennes M\'etropole AES program (project ``SRI''), the European Union's European Regional Development Fund, and the French region of Brittany and Rennes M\'etropole through the contrats de plan \'Etat-R\'egion program (projects ``SOPHIE/STIC \& Ondes'' and ``CyMoCoD'').}
}

\begin{document}
\maketitle

\begin{abstract}
We derive bounds on manipulating partially coherent microwaves
with programmable scattering systems. We first consider the concentration of
power from a partially coherent input into a single output port and derive a
prototype-aware bound by combining multiport network theory (MNT) with
semidefinite relaxation (SDR). We then address general coherency-matrix
synthesis, deriving architecture-independent bounds on fidelity and useful
strength, prototype-specific scalar refinements thereof, and fully
prototype-aware SDR bounds on the useful-strength--fidelity Pareto frontier.
The prototype-aware formulations account for mutual coupling, loss, discrete tunability, and static
scattering. We evaluate the bounds on four
experimental RIS-parametrized MIMO systems with up to 100 1-bit-programmable
elements, using proxy-MNT models estimated from measurements. The resulting
bounds are generally tight compared with feasible discrete-optimization
outcomes. Interestingly, for coherency synthesis, the simpler bounds
can in some cases be tighter than the fully prototype-aware bounds,
highlighting their complementarity. Our results provide certified limits
for wave-domain processing and harvesting of partially coherent
microwaves.
\end{abstract}


\section{Introduction}
\label{sec:introduction}

Coherence describes a wave's ability to interfere with itself. We consider the spatial coherence of single-frequency waves throughout this paper; below, we omit for conciseness the qualifier ``spatial''. Spatial coherence quantifies correlations
between field components associated with distinct spatial points or channels
and thereby determines the stability of the interference produced when these
components are superposed
~\cite{wolf_spaceFrequencyI1982,wolf_spaceFrequencyII1986}. Most theoretical and applied studies of
electromagnetic scattering assume perfectly coherent excitation. Mutual
coherence among independently generated sources generally requires phase
synchronization, which modern multi-antenna communication and sensing systems
routinely implement. Nonetheless, partial coherence or incoherence arises in
many practical settings, including simultaneous unsynchronized transmission
in distributed wireless networks~\cite{vu_noncoherentJT2020,antonioli_mixedCoherentCellFree2023,
ganesan_partiallyCoherentCellFree2024}. Similarly, ambient radio-frequency (RF) energy harvesters may naturally encounter partially coherent or even fully incoherent incident waves~\cite{shen_randomRFmatching2016,shen_multiportRectenna2018,lopez_dynamicRFCombining2022}.

Linear wave--matter interactions generally alter a wave's coherence. The coherence of a wave incident on a scattering system and the latter's scattering matrix fully determine the coherence of the scattered wave. This relation implies general architecture-independent bounds on the concentration of partially coherent or incoherent waves by arbitrary passive scattering systems~\cite{zhang_scatteringConcentration2019,miller2026maximum}. In particular, the single-port scattering-concentration bound (SCB), which upper-bounds the power that a passive linear system can concentrate from the incident wave into a single output port, equals the largest eigenvalue of the incident wave's coherency matrix~\cite{zhang_scatteringConcentration2019}; the SCB is attained when the transmission coefficients from the input ports to the selected output form a unit-norm row vector that is collinear with the conjugate transpose of the dominant eigenvector of the incident wave's coherency matrix.

Recent technological progress in the design and fabrication of programmable metasurfaces, such as reconfigurable intelligent surfaces (RISs) and dynamic metasurface antennas (DMAs), has spurred a plethora of research on electromagnetic scattering in \textit{reconfigurable} wave systems. Existing research,
however, overwhelmingly considers coherently excited reconfigurable wave systems. Moreover, irrespective of the coherence of the system excitation, few results establish fundamental limits on the achievable wave–matter interactions of such systems. Many works are dedicated to optimizing the system's reconfigurable configuration for particular objectives, but these optimizations generally do not identify fundamental limits. Indeed, since the underlying optimization problems are typically high-dimensional, discrete, nonlinear, and nonconvex, scalable optimization methods, in general, do not certify global optimality, and the gap between the best identified configuration and the global optimum remains unknown.

Clearly, a practical reconfigurable wave system cannot realize arbitrary transfer functions. Finite and quantized tunability, attenuation, and uncontrollable scattering by static parts limit a given system’s ability to realize a desired transfer function. Moreover, the effects of the tunable parameters on the transfer function can be highly intertwined due to mutual coupling. Considering the  aforementioned  SCB  for  concreteness, these practical considerations raise the following question: What is the fundamental limit on tuning an experimental prototype of a reconfigurable wave system to concentrate a given incident wave into one output port? More generally, what is the fundamental limit on tuning such a system to implement a desired transformation of the incident wave’s coherence?

At microwave frequencies, programmable wave systems commonly derive their reconfigurability from tunable lumped components such as PIN diodes or varactors. For this reason, they are almost universally amenable to a representation in terms of multiport network theory (MNT) that can account in an electromagnetically consistent manner for all aforementioned practical features (mutual coupling, quantized tunability, attenuation, static parts). Within the MNT framework, each tunable component is represented as a ``virtual'' port terminated by a tunable load. The MNT formulation applies broadly to diverse embodiments of programmable wave systems, including common ones based on RISs and DMAs. Importantly, while the MNT parameters are typically not known a priori for a given prototype, experimental estimation techniques can identify accurate proxy MNT parameters for fabricated prototypes of reconfigurable wave systems
~\cite{sol2024experimentally,ContRIS_LWC,largeRIS_TCOM,del2025experimentalreducedrank,tapie2025experimental,tapie2026channel,del2026cross}. Bounds
evaluated from these parameter sets then incorporate both electromagnetic consistency
and the prototype's actual discrete feasibility set; in other words, these bounds are \textit{prototype-aware}.

Such prototype-aware bounds are useful in two complementary ways. \textit{First,} when a feasible design found by a discrete optimizer approaches the bound, the bound certifies near-optimality: no other admissible configuration of the same prototype can perform substantially better. \textit{Second,} when the bound falls below a desired performance target, it certifies infeasibility: the target cannot be reached with the considered prototype, irrespective of the optimization algorithm used. Bounds therefore provide more than numerical benchmarks; they distinguish algorithmic suboptimality from genuine prototype-level limitations and can guide whether one should improve the optimizer, increase the number of tunable elements, change the hardware, or relax the desired wave transformation.

For coherently excited reconfigurable wave systems, three recent studies derived and evaluated prototype-aware bounds on the achievable SISO channel gain~\cite{salmi2026electromagnetically}, the achievable MIMO aggregate channel gain and operator-synthesis fidelity~\cite{del2026electromagnetic}, and the achievable strength--fidelity tradeoff in output-wavefront synthesis~\cite{del2026prototype}. At the core of these approaches is the formulation of the underlying optimization problem as a quadratically constrained quadratic program (QCQP), or something very similar. A semidefinite relaxation (SDR) of the QCQP then yields a semidefinite program (SDP) that can be solved with standard convex solvers. Because any feasible point of the QCQP is feasible for the SDP, the SDP solution is an upper bound for the QCQP. Closely related theoretical studies of pattern synthesis with reactively loaded arrays, while not grounded in the feasibility set of a fabricated prototype, formulated the problem as a QCQP in~\cite{corcoles2015reactively} and derived SDR-based bounds in~\cite{salmi2025optimization}. More broadly, convex-relaxation-based bounds have been developed for various performance metrics of static electromagnetic structures, including antennas as well as optical and nanophotonic devices~\cite{gustafsson2019tradeoff,angeris2019computational,molesky2020hierarchical,kuang2020computational,gustafsson2020upper,liska2021fundamental,chao2022physical,schab2022upper,angeris2023bounds,shim2024fundamental,gertler2025many,virally2025many,amaolo2026maximum,chao2026blueprints,abdelrahman2026maximumQplanarInductors,miller2026fundamental}. Most of these existing bounds assume coherent excitation; in particular, all
existing prototype-aware bounds for reconfigurable wave systems cited above
do so.

Notwithstanding, combinations of MNT system models with second-order field statistics exist in classical noise-wave analysis~\cite{meys_waveNoise1978,kanaglekar_waveAnalysisNoise1987,wedge_waveTechniquesNoise1992}. More recently, various combinations of MNT system models of reconfigurable wave systems with second-order field statistics were explored for contexts including multiport harvesting of ambient RF fields~\cite{shen_randomRFmatching2016,shen_multiportRectenna2018}, coherency-matrix estimation by multiplexing across a reconfigurable complex system onto a single-port intensity detector~\cite{delHougne_APPAcoherency2025}, and so-called frozen differential scattering\footnote{In frozen differential scattering, a localized perturbation produces a rank-one differential scattering matrix whose
output is coherent irrespective of the input coherence.~\cite{delHougne_frozenDifferential2026}}~\cite{delHougne_frozenDifferential2026}. None of these works provide bounds on specific or generic
coherency transformations. Complementary optical studies considered coherence control with continuously tunable unitary converters
~\cite{guo_unitaryAbsorption2024,guo_unitaryTransmission2024,roquesCarmes_selfConfigCoherence2024}. These results provide important
architecture-independent limits and control strategies, but they do not account for the specificities of the feasibility sets of microwave prototypes of reconfigurable wave systems, where mutual coupling, quantized tunability, attenuation, and static scattering parts matter.

In this paper, we derive and evaluate fundamental limits on transforming the
second-order statistics of partially coherent waves with reconfigurable wave
systems, including both \textit{architecture-independent} and
\textit{prototype-aware} bounds. To the best of our knowledge, the latter are
the first bounds in this setting that explicitly retain the feasibility set of
a specific programmable prototype. Our main contributions are summarized as
follows:
\begin{enumerate}
    \item We derive a \textit{prototype-aware} bound on the maximum power that a
    reconfigurable system can concentrate from a partially coherent input wave
    into a single output port. We formulate the problem as a QCQP, use
    reciprocity to substantially reduce the number of unknowns, and apply an
    SDR to obtain the bound.

    \item We derive \textit{architecture-independent} bounds on transforming a
    given input coherency matrix into a target one. We bound the output's fidelity to the
    target based on the rank of the input's coherency matrix, while we bound its useful strength using passivity.

    \item We derive \textit{prototype-aware} bounds on coherency-matrix synthesis
    with a reconfigurable MIMO system. Using an SDR-certified scaling factor, we
    tighten the architecture-independent useful-strength bound. We then combine a
    lifted SDR with an SDP representation of root fidelity and a Charnes--Cooper
    normalization to bound the useful-strength--fidelity Pareto frontier through
    complementary threshold sweeps.

    \item We evaluate the bounds for four experimental RIS-parametrized MIMO
    systems with up to 100 1-bit-programmable elements and environments ranging
    from rich scattering to free space. We examine the influence of the number
    of programmable elements and the input and target coherency matrices, and
    assess bound tightness against feasible discrete-optimization outcomes.
\end{enumerate}

\textit{Organization:} We introduce our system model in Sec.~\ref{sec:system_model}. We derive our prototype-aware concentration bounds in Sec.~\ref{sec:concentration_bound}. We develop our coherency-synthesis bounds in Sec.~\ref{sec:shaping_pareto_bound}. We summarize our discrete-optimization benchmarks in Sec.~\ref{sec:discrete_optimization_benchmarks}. We present our experimental results in Sec.~\ref{sec_Results}. We briefly conclude in Sec.~\ref{sec:Conclusion}.

\textit{Notation:} \(\mathbf A_{\mathcal B\mathcal C}\) denotes the block of
\(\mathbf A\) selected by row indices \(\mathcal B\) and column indices
\(\mathcal C\). For a Hermitian matrix \(\mathbf A\),
\(\lambda_i(\mathbf A)\) denotes its \(i\)th eigenvalue, indexed in
nonincreasing order, and
\(\lambda_{\max}(\mathbf A)=\lambda_1(\mathbf A)\). The operator
\(\operatorname{vec}(\cdot)\) stacks the columns of a matrix into a vector, and
\(\operatorname{unvec}(\cdot)\) denotes the inverse operation with dimensions
clear from context. \(\operatorname{tr}(\cdot)\) denotes the trace. \((\cdot)^*\), \((\cdot)^\top\), and \((\cdot)^\dagger\) denote complex
conjugation, transpose, and conjugate transpose, respectively.

\section{System Model}
\label{sec:system_model}

We consider a linear, reciprocal, passive, time-invariant wave system with \(N_{\mathrm T}\) input ports, \(N_{\mathrm R}\) distinct output ports, and \(N_{\mathrm S}\) tunable lumped elements. We model each tunable element as a ``virtual'' lumped port terminated by a tunable load.   Denoting by $r_i \in \mathbb{C}$ the reflection coefficient of the $i$th tunable load, we collect the reflection coefficients of all tunable loads in the load vector \(\mathbf r=[r_1,\ldots,r_{N_{\mathrm S}}]^\top \in \mathbb{C}^{N_\mathrm{S}}\). For our experimental prototypes, each tunable load has two possible states whose reflection coefficients are $\alpha\in\mathbb C$ and $\beta\in\mathbb C$, i.e., $r_i\in\{\alpha,\beta\}$. 

Our system can be partitioned into a static subsystem and a tunable subsystem. The tunable subsystem comprises the $N_\mathrm{S}$ tunable loads and is thus characterized by the scattering matrix $\mathbf{\Phi} = \mathrm{diag}(\mathbf{r})\in\mathbb{C}^{N_\mathrm{S}\times N_\mathrm{S}}$. The static subsystem has $N=N_\mathrm{T}+N_\mathrm{R}+N_\mathrm{S}$ ports and is characterized by the scattering matrix $\mathbf{S}\in\mathbb{C}^{N\times N}$; $\mathbf{S}=\mathbf{S}^\top$ since we assume reciprocity. The two subsystems are connected via the $N_\mathrm{S}$ ``virtual'' ports, yielding our overall system. 

The end-to-end transmission matrix $\mathbf{H}\in\mathbb{C}^{N_\mathrm{R}\times N_\mathrm{T}}$ from the $N_\mathrm{T}$ input ports to the $N_\mathrm{R}$ output ports of the overall system maps an incoming power wave $\mathbf{a}\in\mathbb{C}^{N_\mathrm{T}}$ to the associated outgoing power wave $\mathbf{b}\in\mathbb{C}^{N_\mathrm{R}}$:
\begin{equation}
 \mathbf{b}=\mathbf{H}\mathbf{a}.  
 \label{eq_bHa}
\end{equation}
For a given load vector $\mathbf{r}$, according to standard MNT, $\mathbf{H}(\mathbf r)$ is given by
\begin{equation}
  \Hmat(\mathbf r)=
  \mathbf H_0+
  \mathbf A
  \bigl(\mathbf I_{N_{\mathrm S}}-\mathbf\Phi(\mathbf r)\mathbf \Gamma\bigr)^{-1}
  \mathbf\Phi(\mathbf r)\mathbf B,
  \label{eq:mnt_map}
\end{equation}
where, for notational ease, \(\mathbf H_0 \triangleq \mathbf S_{\mathcal R\mathcal T}\in\mathbb C^{N_\mathrm{R}\times N_\mathrm{T}}\), \(\mathbf A \triangleq \mathbf S_{\mathcal R\mathcal S}\in\mathbb C^{N_\mathrm{R}\times N_\mathrm{S}}\), \(\mathbf \Gamma \triangleq \mathbf S_{\mathcal S\mathcal S}\in\mathbb C^{N_\mathrm{S}\times N_\mathrm{S}}\), and \(\mathbf B \triangleq \mathbf S_{\mathcal S\mathcal T}\in\mathbb C^{N_\mathrm{S}\times N_\mathrm{T}}\). Here \(\mathcal T\), \(\mathcal R\), and \(\mathcal S\) denote the index sets associated with the input ports, output ports, and ``virtual'' ports, respectively. 
We emphasize that the described MNT-based system model accounts for all static scattering (including structural scattering and environmental scattering) and makes no assumptions about the distances between the involved entities (in particular, no far-field assumption is required).

The MNT-based model parameters
$\{\alpha,\beta,\mathbf{H}_0,\mathbf{A},\mathbf{\Gamma},\mathbf{B}\}$
are usually not known a priori for an experimental prototype. If the system were perfectly specified, the MNT parameters could in principle be extracted from component data sheets and full-wave simulations~\cite{tapie2023systematic}; in practice, however, fabrication tolerances, unknown environmental details, incompletely documented components, proprietary designs, and limited computational resources often preclude such an approach. Moreover, these MNT parameters cannot be measured directly on the experimental prototype because the tunable elements are typically not connectorized and their number far exceeds the number of available ports of a typical vector network analyzer (VNA). One can nonetheless estimate proxy MNT parameter sets that are operationally equivalent, in the sense that they accurately predict the physically measurable end-to-end scattering parameters for any admissible configuration of the tunable loads~\cite{sol2024experimentally,del2025physics,ContRIS_LWC,largeRIS_TCOM,del2025experimentalreducedrank,tapie2025experimental,tapie2026channel,del2026cross}. These proxy MNT parameters are not unique: they are related by gauge freedoms such as diagonal-similarity, complex-scaling, and M\"obius transformations, as detailed in the appendices of~\cite{salmi2026electromagnetically,del2026electromagnetic,del2026prototype}. As in~\cite{salmi2026electromagnetically,del2026electromagnetic,
del2026prototype}, the bounds presented below are invariant under
reciprocity-preserving changes of proxy MNT parameter sets.

A partially coherent power wave is characterized by its second-order statistics captured by a coherency matrix. For the input and output waves relevant to our scenario, the associated coherency matrices are
\begin{subequations}
\begin{equation}
\Cin=\mathbb E[\mathbf a\mathbf a^\dagger],
\end{equation}
\begin{equation}
\Cout=\mathbb E[\mathbf b\mathbf b^\dagger], 
\end{equation}
\end{subequations}
and their relation is fixed by $\mathbf{H}(\mathbf{r})$:
\begin{equation}
  \Cout(\mathbf r)
  =
  \Hmat(\mathbf r)\,\Cin\,\Hmat^\dagger(\mathbf r).
  \label{eq:coherency_propagation}
\end{equation}

\section{Concentration Bound}
\label{sec:concentration_bound}

Before addressing in Sec.~\ref{sec:shaping_pareto_bound} the general problem of which transformations of a coherency matrix are achievable with a given prototype of a reconfigurable wave system, we first study the important special case of maximally concentrating a partially coherent input wave into one output port with a given prototype. This problem, where \(N_\mathrm{T}>1\) and \(N_\mathrm{R}=1\), appears whenever one seeks to collect power from spatially incoherent or partially coherent waves.

If these waves interact with a lossless system that can realize an arbitrary unitary transformation, the maximum power that can be concentrated into one output port is
\begin{equation}
    P_{\mathrm{SCB}}=\lambda_{\max}(\Cin),
    \label{eq:scattering_concentration_bound}
\end{equation}
where \(\lambda_{\max}(\Cin)\) denotes the largest eigenvalue of the input coherency matrix~\cite{zhang_scatteringConcentration2019}. While this SCB is a useful architecture-independent benchmark, it does not account for the restrictions of a concrete programmable wave system. In particular, the reconfigurable wave systems we consider experimentally are generally lossy and cannot realize arbitrary unitary transformations. This motivates our prototype-aware SCB.

\subsection{Problem Formulation and Reciprocity}
We seek a prototype-aware bound for the single-port concentrated power
\begin{equation}
    P_1(\mathbf r)
    =
    \mathbf h(\mathbf r)\Cin\mathbf h^\dagger(\mathbf r),
    \label{eq:raw_concentrated_power}
\end{equation}
where \(\mathbf h(\mathbf r)\in\mathbb C^{1\times N_\mathrm T}\) denotes the only row of \(\Hmat(\mathbf r)\) for \(N_\mathrm R=1\).
Formally, the concentration problem is
\begin{equation}
\begin{aligned}
    \max_{\mathbf r}\quad
    & \mathbf h(\mathbf r)\Cin\mathbf h^\dagger(\mathbf r) \\
    \mathrm{s.t.}\quad
    & r_i\in\{\alpha,\beta\},\qquad i=1,\ldots,N_\mathrm{S}.
\end{aligned}
\label{eq:concentration_problem}
\end{equation}

A direct treatment of \eqref{eq:concentration_problem} is possible but inconvenient. The reason relates to the system's multiple-input nature. Analogously to previous work on a multi-input multi-output (MIMO) system in~\cite{del2026electromagnetic}, we would have to introduce an auxiliary variable of dimensions $N_\mathrm{S} \times N_\mathrm{T}$ and then constrain its rows with dedicated repetition constraints to correspond to the same load states. Since our present problem is a multiple-input single-output (MISO) problem for a reciprocal system, we can avoid this complication (unlike the MIMO case in~\cite{del2026electromagnetic}) by considering the reciprocal single-input multiple-output (SIMO) problem. With a single input, the required auxiliary variable is an $N_\mathrm{S}$-element vector (i.e., the optimization variable has $N_\mathrm{T}$ times fewer entries) and we do not need any repetition constraints. Related reciprocity-based tricks were recently leveraged in some works on photonic inverse design~\cite{yao_traceIncoherent2022,pestourie_incoherentMetasurfaces2023,roquesCarmes_scintillation2022}.

Since we assume that our system is reciprocal, the end-to-end transmission matrix $\mathbf{G}\in\mathbb{C}^{N_\mathrm{T}\times N_\mathrm{R}}$ from the $N_\mathrm{R}$ receive ports to the $N_\mathrm{T}$ input ports of the overall system is
\begin{equation}
  \mathbf{G}(\mathbf r) \triangleq  \left(\Hmat(\mathbf r)\right)^\top =
    \mathbf H_0^\top+
  \mathbf B^\top
  \bigl(\mathbf I_{N_{\mathrm S}}-\mathbf\Phi(\mathbf r)\mathbf \Gamma\bigr)^{-1}
  \mathbf\Phi(\mathbf r)\mathbf A^\top,
  \label{eq:mnt_map_reciprocal}
\end{equation}
where we used \(\mathbf\Gamma=\mathbf\Gamma^\top\), \(\mathbf\Phi=\mathbf\Phi^\top\), and \(\mathbf\Phi(\mathbf I-\mathbf\Gamma\mathbf\Phi)^{-1}
=(\mathbf I-\mathbf\Phi\mathbf\Gamma)^{-1}\mathbf\Phi\).
Specializing the notation to our case with $N_\mathrm{R}=1$, we have
\begin{equation}
    \mathbf g(\mathbf r) \triangleq \left(\mathbf h(\mathbf r)\right)^\top\in\mathbb C^{N_\mathrm{T}}.
    \label{eq:h_g_reciprocity}
\end{equation}
Substituting \eqref{eq:h_g_reciprocity} into \eqref{eq:raw_concentrated_power} yields
\begin{equation}
    P_1(\mathbf r)
    =
    \mathbf g^\dagger(\mathbf r)\Cin^\top\mathbf g(\mathbf r),
    \label{eq:reciprocal_concentration_objective}
\end{equation}
and substituting \eqref{eq:reciprocal_concentration_objective} into~\eqref{eq:concentration_problem} yields
\begin{equation}
\begin{aligned}
    \max_{\mathbf r}\quad
    & \mathbf g^\dagger(\mathbf r)\Cin^\top\mathbf g(\mathbf r) \\
    \mathrm{s.t.}\quad
    & r_i\in\{\alpha,\beta\},\qquad i=1,\ldots,N_\mathrm{S}.
\end{aligned}
\label{eq:reciprocal_concentration_problem}
\end{equation}

\subsection{QCQP Formulation}
To formulate~\eqref{eq:reciprocal_concentration_problem} as a QCQP, we introduce the auxiliary variable
\begin{equation}
    \mathbf y(\mathbf r)
    \triangleq
    \bigl(\mathbf I_{N_\mathrm{S}}-\mathbf\Phi(\mathbf r)\mathbf\Gamma\bigr)^{-1}
    \mathbf\Phi(\mathbf r)\mathbf A^\top
    \in\mathbb C^{N_\mathrm{S}}.
    \label{eq:reciprocal_auxiliary_variable}
\end{equation}
Using~\eqref{eq:reciprocal_auxiliary_variable}, we can rewrite
\eqref{eq:mnt_map_reciprocal}, specialized to \(N_\mathrm R=1\), as
\begin{equation}
    \mathbf g
    =
    \mathbf H_0^\top+\mathbf B^\top\mathbf y,
    \label{eq:reciprocal_response_internal_state}
\end{equation}
and~\eqref{eq:reciprocal_concentration_objective} as
\begin{equation}
    P_1
    =
    \left(
    \mathbf H_0^\top+\mathbf B^\top\mathbf y
    \right)^\dagger
    \Cin^\top
    \left(
    \mathbf H_0^\top+\mathbf B^\top\mathbf y
    \right),
    \label{eq:concentration_quadratic_objective}
\end{equation}
which is a quadratic formulation of our objective with respect to \(\mathbf y\).

We are left with imposing that \(\mathbf y\) is compatible with feasible binary load vectors. From~\eqref{eq:reciprocal_auxiliary_variable}, \(\mathbf y\) equivalently satisfies
\begin{equation}
    \mathbf y
    =
    \mathbf\Phi(\mathbf r)
    \left(
    \mathbf\Gamma\mathbf y+\mathbf A^\top
    \right).
    \label{eq:reciprocal_auxiliary_fixed_point}
\end{equation}
Denoting by \(\boldsymbol\gamma_i^\top\) the \(i\)th row of \(\mathbf\Gamma\), and by \(a_i\) the \(i\)th entry of \(\mathbf A^\top\) ($\mathbf{A}$ is a vector for $N_\mathrm{R}=1$), the \(i\)th entry of~\eqref{eq:reciprocal_auxiliary_fixed_point} reads
\begin{equation}
    y_i
    =
    r_i
    \left(
    \boldsymbol\gamma_i^\top\mathbf y+a_i
    \right).
    \label{eq:yi_binary_load_relation}
\end{equation}
As in~\cite{shim2024fundamental,gertler2025many,salmi2026electromagnetically,del2026electromagnetic,del2026prototype}, we can now express our binary-programmability constraint \(r_i\in\{\alpha,\beta\}\) based on~\eqref{eq:yi_binary_load_relation} as 
\begin{equation}
\begin{aligned}
    &
    \left[
    y_i-\alpha u_i(\mathbf y)
    \right]^*
    \left[
    y_i-\beta u_i(\mathbf y)
    \right]
    =0,\\
    &\hfill i=1,\ldots,N_\mathrm{S},
\end{aligned}
\label{eq:binary_quadratic_concentration}
\end{equation}
where $u_i(\mathbf y)\triangleq \boldsymbol\gamma_i^\top\mathbf y+a_i$.

Combining~\eqref{eq:concentration_quadratic_objective} and~\eqref{eq:binary_quadratic_concentration}, we obtain the QCQP formulation of~\eqref{eq:reciprocal_concentration_problem}:
\begin{equation}
\begin{aligned}
    \max_{\mathbf y}\quad
    &
    \left(
    \mathbf H_0^\top+\mathbf B^\top\mathbf y
    \right)^\dagger
    \Cin^\top
    \left(
    \mathbf H_0^\top+\mathbf B^\top\mathbf y
    \right) \\
    \mathrm{s.t.}\quad
    &
    \left[
    y_i-\alpha
    \left(
    \boldsymbol\gamma_i^\top\mathbf y+a_i
    \right)
    \right]^*
    \left[
    y_i-\beta
    \left(
    \boldsymbol\gamma_i^\top\mathbf y+a_i
    \right)
    \right]
    =
    0,\\
    &\hfill i=1,\ldots,N_\mathrm{S}.
\end{aligned}
\label{eq:concentration_qcqp}
\end{equation}

Defining \(\mathbf R_0\triangleq \mathbf B^*\Cin^\top\mathbf B^\top\), \(\mathbf q_0\triangleq \mathbf B^*\Cin^\top\mathbf H_0^\top\), and \(t_0\triangleq \left(\mathbf H_0^\top\right)^\dagger\Cin^\top\mathbf H_0^\top\), we can rewrite our objective as
\begin{equation}
    P_1
    =
    \mathbf y^\dagger\mathbf R_0\mathbf y
    +
    2\mathrm{Re}\{\mathbf q_0^\dagger\mathbf y\}
    +
    t_0 .
    \label{eq:concentration_expanded_objective}
\end{equation}
Similarly, defining \(\mathbf p_{\alpha,i}\triangleq\mathbf e_i-\alpha\boldsymbol\gamma_i\), \(\mathbf p_{\beta,i}\triangleq\mathbf e_i-\beta\boldsymbol\gamma_i\), \(d_{\alpha,i}\triangleq-\alpha a_i\), \(d_{\beta,i}\triangleq-\beta a_i\), \(\mathbf R_i\triangleq\mathbf p_{\alpha,i}^*\mathbf p_{\beta,i}^\top\), \(\mathbf q_{1,i}\triangleq d_{\beta,i}\mathbf p_{\alpha,i}^*\), \(\mathbf q_{2,i}\triangleq d_{\alpha,i}\mathbf p_{\beta,i}^*\), and \(t_i\triangleq d_{\alpha,i}^*d_{\beta,i}\), where \(\mathbf e_i\) denotes the \(i\)th canonical basis vector of \(\mathbb C^{N_\mathrm S}\), we can rewrite the \(i\)th binary-programmability constraint as
\begin{equation}
    \mathbf y^\dagger\mathbf R_i\mathbf y
    +
    \mathbf y^\dagger\mathbf q_{1,i}
    +
    \mathbf q_{2,i}^\dagger\mathbf y
    +
    t_i
    =
    0.
    \label{eq:concentration_expanded_constraint}
\end{equation}
The matrices \(\mathbf R_i\) are generally non-Hermitian, so we impose the real and imaginary parts of~\eqref{eq:concentration_expanded_constraint} separately.
Combining~\eqref{eq:concentration_expanded_objective} and~\eqref{eq:concentration_expanded_constraint}, we can rewrite the QCQP in~\eqref{eq:concentration_qcqp} as
\begin{equation}
\begin{aligned}
    \max_{\mathbf y}\quad
    &
    \mathbf y^\dagger\mathbf R_0\mathbf y
    +
    2\mathrm{Re}\{\mathbf q_0^\dagger\mathbf y\}
    +
    t_0\\
    \mathrm{s.t.}\quad
    &
    \mathbf y^\dagger\mathbf R_i\mathbf y
    +
    \mathbf y^\dagger\mathbf q_{1,i}
    +
    \mathbf q_{2,i}^\dagger\mathbf y
    +
    t_i
    =
    0,\quad i=1,\ldots,N_\mathrm S .
\end{aligned}
\label{eq:concentration_qcqp_compact}
\end{equation}

\subsection{SDR Bound}
We now lift the quadratic terms by defining \(\mathbf Y\triangleq\mathbf y\mathbf y^\dagger\). Using the identity \(\mathbf y^\dagger\mathbf R_i\mathbf y=\mathrm{tr}(\mathbf R_i\mathbf y\mathbf y^\dagger)=\mathrm{tr}(\mathbf R_i\mathbf Y)\), we can rewrite the QCQP in~\eqref{eq:concentration_qcqp_compact} as
\begin{equation}
\begin{aligned}
    \max_{\mathbf Y,\mathbf y}\quad
    &
    \mathrm{tr}(\mathbf R_0\mathbf Y)
    +
    2\mathrm{Re}\{\mathbf q_0^\dagger\mathbf y\}
    +
    t_0\\
    \mathrm{s.t.}\quad
    &
    \mathrm{tr}(\mathbf R_i\mathbf Y)
    +
    \mathbf y^\dagger\mathbf q_{1,i}
    +
    \mathbf q_{2,i}^\dagger\mathbf y
    +
    t_i=0,\quad i=1,\ldots,N_\mathrm S,\\
    &
    \mathbf Y=\mathbf y\mathbf y^\dagger .
\end{aligned}
\label{eq:concentration_lifted_qcqp}
\end{equation}
The only non-convex constraint in~\eqref{eq:concentration_lifted_qcqp} is \(\mathbf Y=\mathbf y\mathbf y^\dagger\). Following standard SDR arguments~\cite{luo2010semidefinite,boyd2004convex}, we relax the non-convex equality \(\mathbf Y=\mathbf y\mathbf y^\dagger\) to the convex inequality \(\mathbf Y\succeq\mathbf y\mathbf y^\dagger\) which can equivalently be expressed as 
\(\left[
\begin{smallmatrix}
\mathbf Y & \mathbf y\\
\mathbf y^\dagger & 1
\end{smallmatrix}
\right]\succeq\mathbf 0\).
This relaxation yields the SDP
\begin{equation}
\begin{aligned}
    \max_{\mathbf Y,\mathbf y}\quad
    &
    \mathrm{tr}(\mathbf R_0\mathbf Y)
    +
    2\mathrm{Re}\{\mathbf q_0^\dagger\mathbf y\}
    +
    t_0\\
    \mathrm{s.t.}\quad
    &
    \mathrm{tr}(\mathbf R_i\mathbf Y)
    +
    \mathbf y^\dagger\mathbf q_{1,i}
    +
    \mathbf q_{2,i}^\dagger\mathbf y
    +
    t_i=0,\quad i=1,\ldots,N_\mathrm S,\\
    &
    \begin{bmatrix}
        \mathbf Y & \mathbf y\\
        \mathbf y^\dagger & 1
    \end{bmatrix}
    \succeq \mathbf 0 .
\end{aligned}
\label{eq:concentration_sdr}
\end{equation}
Again, the generally complex-valued equality constraints are independently imposed for their real and imaginary parts. The SDP in~\eqref{eq:concentration_sdr} is convex and can be solved with standard SDP solvers. Since every feasible binary configuration yields a feasible rank-one point of~\eqref{eq:concentration_sdr}, whereas the SDR also allows higher-rank matrices, the optimal value of~\eqref{eq:concentration_sdr} upper-bounds the achievable concentration.

By replacing \(\Cin^\top\) in the objective of \eqref{eq:concentration_sdr} with \(\mathbf I_{N_\mathrm T}\), we can further obtain an upper bound on the aggregate channel gain $\|\mathbf h(\mathbf r)\|_2^2$ from the $N_\mathrm{T}$ input ports to the output port. We denote this bound by $\bar\eta_{\mathrm c}$, such that $\max_{\mathbf r}\|\mathbf h(\mathbf r)\|_2^2
    \leq \bar\eta_{\mathrm c}$. Combining this bound with passivity and
\(\Cin\preceq\lambda_{\max}(\Cin)\mathbf I_{N_\mathrm T}\), we obtain
\begin{equation}
    P_1
    \leq
    \min\{\bar\eta_{\mathrm c},1\}\lambda_{\max}(\Cin)
    =
    \min\{\bar\eta_{\mathrm c},1\}P_{\mathrm{SCB}} .
    \label{eq:transmission_scaled_scb}
\end{equation}
This bound complements the direct SDR bound from~\eqref{eq:concentration_sdr}.

Within the reciprocal proxy-MNT representation used in this work, the two
SDR-based bounds derived in this section are insensitive to
reciprocity-preserving proxy-MNT ambiguities for the same reason as in the
ambiguity-insensitivity appendix of~\cite{salmi2026electromagnetically}. In
short, these ambiguities amount to invertible changes of internal coordinates
that leave the physical response unchanged for every admissible load vector.
Since both objectives depend on the proxy model only through the physical
response \(\mathbf h\), the corresponding change of variables maps feasible
SDP points to feasible SDP points with the same objective values. Hence, both
bounds are invariant under these reciprocity-preserving ambiguities.

\section{Coherency-Synthesis Bounds}
\label{sec:shaping_pareto_bound}

We now consider the more general task of synthesizing a desired output coherency matrix rather than only concentrating power into one output port; thus, unlike Sec.~\ref{sec:concentration_bound}, we do \textit{not} limit ourselves to \(N_\mathrm R=1\) in this section and consider a generic MIMO system with \(N_\mathrm R>1\) instead.  We assume that the input coherency matrix \(\Cin\) is fixed and known. 
Our goal is for the realized output coherency matrix defined in~\eqref{eq:coherency_propagation} to resemble as closely as possible a desired target coherency matrix \(\rhostar\succeq\mathbf 0\), normalized such that \(\mathrm{tr}(\rhostar)=1\).

\subsection{Performance Metrics}

For practical coherency-matrix synthesis, matching the normalized target
structure is not sufficient: the system must also deliver appreciable power
in a form compatible with that structure. We therefore consider two metrics:
\textit{fidelity} and \textit{useful strength}. For nonzero output power, we
first normalize the output coherency matrix as
\begin{equation}
    \rhoout(\mathbf r)
    =
    \frac{\Cout(\mathbf r)}
    {\operatorname{tr}\!\left(\Cout(\mathbf r)\right)} .
    \label{eq:normalized_output_coherency}
\end{equation}
We quantify the \textit{fidelity}, i.e., the structural agreement between
\(\rhoout(\mathbf r)\) and \(\rhostar\), using the squared Uhlmann--Jozsa
fidelity~\cite{uhlmann1976transition,jozsa1994fidelity},
\begin{equation}
    F(\mathbf r)
    =
    \left[
    \operatorname{tr}\sqrt{
    \sqrt{\rhostar}\,
    \rhoout(\mathbf r)\,
    \sqrt{\rhostar}}
    \right]^2 ,
    \label{eq:coherency_fidelity}
\end{equation}
which satisfies \(0\leq F(\mathbf r)\leq1\) and is insensitive to the
overall output-power scale.
We further define the \textit{useful strength} as the total output power
weighted by its fidelity to the desired coherency structure:
\begin{equation}
\begin{aligned}
    U(\mathbf r)
    =
    \operatorname{tr}\!\left(\Cout(\mathbf r)\right)F(\mathbf r)
    =
    \left[
    \operatorname{tr}\sqrt{
    \sqrt{\rhostar}\,
    \Cout(\mathbf r)\,
    \sqrt{\rhostar}}
    \right]^2 .
\end{aligned}
\label{eq:coherency_useful_strength}
\end{equation}
Unlike the total output power \(\operatorname{tr}(\Cout)\), \(U\) discounts
power associated with an output coherency structure that differs from the
target.

The scale-invariant role of \(F\) parallels that of fidelity metrics in
previous bounds for MIMO-operator synthesis~\cite{del2026electromagnetic}
and output-wavefront synthesis~\cite{del2026prototype} with coherently
excited programmable wave systems,\footnote{More broadly, normalized efficiency metrics and their
bounds also play an important role in photonic
design~\cite{schab2022upper,angeris2023bounds}.} while \(U\) generalizes the target-mode
strength metric of~\cite{del2026prototype}. Indeed, for
\(\Cout=\boldsymbol\psi\boldsymbol\psi^\dagger\) and
\(\rhostar=\hat{\boldsymbol\psi}_\star
\hat{\boldsymbol\psi}_\star^\dagger\), with
\(\lVert\hat{\boldsymbol\psi}_\star\rVert_2=1\), the two metrics reduce to
\(
F=\lvert\hat{\boldsymbol\psi}_\star^\dagger
\boldsymbol\psi\rvert^2 / \lVert\boldsymbol\psi\rVert_2^2 \) and \(U=
\lvert\hat{\boldsymbol\psi}_\star^\dagger
\boldsymbol\psi\rvert^2
\).
Thus, when both \(\Cin\) and \(\rhostar\) have rank one, they coincide with the normalized output-wavefront
overlap and target-mode power, respectively, used
in~\cite{del2026prototype}.

\subsection{Architecture-Independent Bounds}

Before accounting for the restrictions of a particular programmable
prototype, deterministic linear scattering already imposes fundamental
constraints on coherency synthesis.
Since multiplication by a deterministic linear operator cannot increase the matrix rank, it follows from \eqref{eq:coherency_propagation} that
\begin{equation}
    \operatorname{rank}(\Cout)
    \leq
    \operatorname{rank}(\Cin).
\end{equation}
Consequently, exact nonzero synthesis of an output proportional to
\(\rhostar\) requires
\(
    \operatorname{rank}(\rhostar)
    \leq
    \operatorname{rank}(\Cin).
\)

More generally, any output coherency matrix has rank at most
\(r\triangleq\min\{\operatorname{rank}(\Cin),N_{\mathrm R}\}\).
Let \(\mathbf P_{\mathrm{out}}\) denote the orthogonal projector onto its support.
Since \(\rhoout\) has no support outside this subspace,
\(F\leq\operatorname{tr}(\mathbf P_{\mathrm{out}}\rhostar)\).
This follows by combining the equivalent norm representation of fidelity, \(F=\|\rhoout^{1/2}\rhostar^{1/2}\|_1^2\)~\cite{watrous2018theory}, with H\"older's inequality for Schatten norms~\cite{bhatia1997matrix}.
By the Ky Fan maximum principle~\cite{fan1949weyl},
\(\operatorname{tr}(\mathbf P_{\mathrm{out}}\rhostar)\) is maximized over all
\(r\)-dimensional subspaces by the \(r\) dominant eigenvectors of \(\rhostar\), yielding the architecture-independent fidelity bound
\begin{equation}
    F
    \leq
    F_{\mathrm{rank}}
    \triangleq
    \sum_{i=1}^{r}\lambda_i(\rhostar).
    \label{eq:rank_fidelity_bound}
\end{equation}
This bound constrains the achievable fidelity irrespective of useful
strength.

Passivity imposes a complementary architecture-independent limit on useful
strength. From
\eqref{eq:coherency_useful_strength} and the singular-value inequality for
matrix products~\cite[Th.~3.3.14(a)]{horn1991topics},
\begin{equation}
    \sqrt{U}
    \leq
    \sum_{i=1}^{M}
    \sqrt{\lambda_i(\Cout)\lambda_i(\rhostar)} ,
    \label{eq:useful_strength_spectral_bound}
\end{equation}
where \(M\triangleq\min(N_{\mathrm T},N_{\mathrm R})\).
Combining this inequality with the maximum-brightness theorem for passive
linear scattering,
\(\lambda_i(\Cout)\leq\lambda_i(\Cin)\)~\cite{miller2026maximum}, yields
\begin{equation}
    U
    \leq
    U_{\mathrm{MB}}
    \triangleq
    \left[
        \sum_{i=1}^{M}
        \sqrt{\lambda_i(\Cin)\lambda_i(\rhostar)}
    \right]^2 .
    \label{eq:maximum_brightness_useful_strength}
\end{equation}
This bound holds irrespective of output fidelity and reduces to the
scattering-concentration bound in
\eqref{eq:scattering_concentration_bound} when \(\rhostar\) has rank one.\footnote{At exact synthesis (\(F=1\)), \(\Cout=U\rhostar\), so the
maximum-brightness theorem~\cite{miller2026maximum} further implies
\(
U\leq
\min_{1\leq i\leq\operatorname{rank}(\rhostar)}
\lambda_i(\Cin)/\lambda_i(\rhostar).
\)}

The maximum-brightness bound on useful strength in
\eqref{eq:maximum_brightness_useful_strength} can be tightened further by
accounting for a given prototype's maximum output power under a unit-power
coherent input. We denote by \(\bar\eta_{\mathrm{tr}}\) a bound on this quantity, such that
\(\max_{\mathbf r,\,\|\mathbf q\|_2=1}
\|\Hmat(\mathbf r)\mathbf q\|_2^2
\leq\bar\eta_{\mathrm{tr}}\).
A certified \(\bar\eta_{\mathrm{tr}}\) can be obtained by applying the joint
input-wavefront-and-configuration SDR framework of~\cite{del2026prototype}
with total output power as the objective, and taking the minimum of the
resulting bound and the passivity limit \(1\).
Since \(\Hmat^\dagger\Hmat\preceq\bar\eta_{\mathrm{tr}}\mathbf I\), the
nonzero eigenvalues of \(\Cout=\Hmat\Cin\Hmat^\dagger\) equal those of
\(\Cin^{1/2}\Hmat^\dagger\Hmat\Cin^{1/2}
\preceq\bar\eta_{\mathrm{tr}}\Cin\). Hence,
\begin{equation}
    \lambda_i(\Cout)\leq
    \bar\eta_{\mathrm{tr}}\lambda_i(\Cin),
    \qquad i=1,\ldots,M.
    \label{eq:prototype_scaled_eigenvalues}
\end{equation}
Substituting~\eqref{eq:prototype_scaled_eigenvalues} into
\eqref{eq:useful_strength_spectral_bound} yields a scaled maximum-brightness
bound on useful strength:
\begin{equation}
    U\leq U_{\mathrm{SMB}}
    \triangleq\bar\eta_{\mathrm{tr}}U_{\mathrm{MB}} .
    \label{eq:transmission_scaled_maximum_brightness}
\end{equation}
Unlike \(U_{\mathrm{MB}}\), \(U_{\mathrm{SMB}}\) is a prototype-aware bound,
but it retains only a scalar power-gain ceiling. We next derive the full
prototype-aware bounds that additionally retain how the feasible transmission
responses relate to the specific input and target coherency matrices.

\subsection{Coherent-Mode Representation}

For a compact formulation of the prototype-feasibility constraints, we work in the
coherent-mode basis of \(\Cin\) rather than in the physical input-port basis,
based on the standard coherent-mode representation of partially coherent
fields~\cite{wolf_spaceFrequencyI1982,wolf_spaceFrequencyII1986,withington_modalPartialCoherence1998,zhang_scatteringConcentration2019,roquesCarmes_selfConfigCoherence2024,
guo_unitaryAbsorption2024,miller2026fundamental}:
\begin{equation}
    \Cin
    =
    \mathbf U_K\boldsymbol\Lambda_K\mathbf U_K^\dagger
    =
    \mathbf L\mathbf L^\dagger,
    \label{eq:coherent_mode_decomposition}
\end{equation}
where \(K=\operatorname{rank}(\Cin)\),
\(\mathbf U_K\in\mathbb C^{N_\mathrm T\times K}\) contains the eigenvectors
associated with the nonzero eigenvalues of \(\Cin\),
\(\boldsymbol\Lambda_K\in\mathbb R_+^{K\times K}\) is the corresponding
diagonal matrix of eigenvalues, and
\(\mathbf L\triangleq
\mathbf U_K\boldsymbol\Lambda_K^{1/2}
\in\mathbb C^{N_\mathrm T\times K}\).
The columns of \(\mathbf U_K\) are mutually orthogonal coherent modes whose
modal amplitudes are mutually uncorrelated, while the corresponding
eigenvalues give their average powers.

Compared to a direct formulation in the physical input-port basis, this
basis choice offers two benefits. \textit{First}, the diagonal modal
statistics allow the output coherency matrix to be expressed as a sum of
modal output coherencies without cross terms between distinct modes:
\begin{equation}
    \Cout
    =
    \sum_{k=1}^{K}
    \left(\Hmat(\mathbf r)\mathbf l_k\right)
    \left(\Hmat(\mathbf r)\mathbf l_k\right)^\dagger,
    \label{eq:coherency_modal_sum}
\end{equation}
where \(\mathbf l_k\) denotes the \(k\)th column of \(\mathbf L\), i.e., the
\(k\)th power-weighted coherent-mode vector.
\textit{Second}, only the \(K=\operatorname{rank}(\Cin)\) excited modes need
to be represented, rather than all \(N_\mathrm T\) input-port excitations in
a straightforward port-basis formulation, reducing the number of auxiliary
variables whenever \(\Cin\) is rank deficient.

To express the modal responses in terms of the MNT model, we define
\(\mathbf H_\mathrm e\triangleq\mathbf H_0\mathbf L
\in\mathbb C^{N_\mathrm R\times K}\) and
\(\mathbf B_\mathrm e\triangleq\mathbf B\mathbf L
\in\mathbb C^{N_\mathrm S\times K}\).
Substituting~\eqref{eq:mnt_map} into~\eqref{eq:coherency_modal_sum}, we obtain
\begin{equation}
    \Cout(\mathbf r)
    =
    \left(\mathbf H_\mathrm e+\mathbf A\mathbf X(\mathbf r)\right)
    \left(\mathbf H_\mathrm e+\mathbf A\mathbf X(\mathbf r)\right)^\dagger,
    \label{eq:coherency_mode_factorization}
\end{equation}
where we define the auxiliary matrix-valued variable
\begin{equation}
    \mathbf X(\mathbf r)
    \triangleq
    \left(\mathbf I_{N_\mathrm S}
    -\mathbf\Phi(\mathbf r)\mathbf\Gamma\right)^{-1}
    \mathbf\Phi(\mathbf r)\mathbf B_\mathrm e
    \in\mathbb C^{N_\mathrm S\times K}.
    \label{eq:coherency_internal_matrix}
\end{equation}
The \(k\)th column \(\mathbf x_k\) of \(\mathbf X\) is therefore the auxiliary vector-valued variable associated with the \(k\)th coherent-mode excitation. We likewise denote the \(k\)th columns of \(\mathbf H_\mathrm e\) and \(\mathbf B_\mathrm e\) by \(\mathbf h_k\) and \(\mathbf b_k\), respectively, and define
\(\mathbf x\triangleq\operatorname{vec}(\mathbf X)\in\mathbb C^{N_\mathrm S  K}\).

\subsection{Prototype-Feasibility Constraints}
Defining \(u_{i,k}\triangleq\boldsymbol\gamma_i^\top\mathbf x_k+b_{i,k}\), the binary-programmability constraint for element \(i\) and mode \(k\) follows directly from~\eqref{eq:binary_quadratic_concentration} as
\begin{equation}
    \begin{aligned}[t]
     &\left[x_{i,k}-\alpha u_{i,k}\right]^*
    \left[x_{i,k}-\beta u_{i,k}\right]=0,
    \\
        &i=1,\ldots,N_\mathrm S,\\
        &k=1,\ldots,K.
    \end{aligned}
    \label{eq:coherency_per_mode_binary}
\end{equation}

For \(K=1\), corresponding to a fully coherent input, the per-mode constraints
in~\eqref{eq:coherency_per_mode_binary} are sufficient. For \(K>1\), because all \(K\) modal responses are produced by the same physical load vector
\(\mathbf r\), the per-mode constraints in~\eqref{eq:coherency_per_mode_binary}
alone are insufficient: they allow different coherent modes to select different
states of the same tunable element. We therefore additionally impose repetition constraints that enforce one common state of every tunable element across all coherent
modes. Such repetition constraints are commonly needed when several excitations must
correspond to the same physical system state~\cite{shim2024fundamental,gertler2025many,salmi2025optimization,del2026electromagnetic}. They closely resemble the repetition constraints used for coherent MIMO
operator synthesis in~\cite{del2026electromagnetic}, with the distinction that
the excitation columns here correspond to coherent modes of \(\Cin\).

To couple the load-state choices across modes, we consider two alternative repetition-constraint formulations: a compact single-reference formulation and a stricter all-pairs formulation. For the compact single-reference formulation, we choose \(k_0=1\) and impose
\begin{subequations}
\label{eq:coherency_repetition_constraints}
\begin{align}
    &\left[x_{i,k}-\alpha u_{i,k}\right]^*
    \left[x_{i,k_0}-\beta u_{i,k_0}\right]=0,
    \\
    &\left[x_{i,k}-\beta u_{i,k}\right]^*
    \left[x_{i,k_0}-\alpha u_{i,k_0}\right]=0,
    \\
    &i=1,\ldots,N_\mathrm S,
    \notag\\
    &k=1,\ldots,K,\qquad k\ne k_0.
    \notag
\end{align}
\end{subequations}
This single-reference formulation is compact, adding only \(2N_\mathrm S(K-1)\) complex equalities.
Together with the per-mode binary-programmability constraints in~\eqref{eq:coherency_per_mode_binary}, the single-reference constraints in~\eqref{eq:coherency_repetition_constraints} enforce a common load state between the reference mode and every other mode with
\(
|x_{i,k_0}|+|u_{i,k_0}|>0.
\)
If
\(
x_{i,k_0}=u_{i,k_0}=0,
\)
the reference mode does not constrain the load state of element \(i\), and different nonzero modes could still select different states.

To avoid this reference-mode dependence, we also consider an alternative all-pairs formulation in which, for every unordered pair of distinct modes \(1\leq k<\ell\leq K\), we impose
\begin{subequations}
\label{eq:coherency_repetition_constraints_all_pairs}
\begin{align}
    &\left[x_{i,k}-\alpha u_{i,k}\right]^*
    \left[x_{i,\ell}-\beta u_{i,\ell}\right]=0,
    \\
    &\left[x_{i,k}-\beta u_{i,k}\right]^*
    \left[x_{i,\ell}-\alpha u_{i,\ell}\right]=0,
    \\
    &i=1,\ldots,N_\mathrm S,
    \notag\\
    &1\leq k<\ell\leq K.
    \notag
\end{align}
\end{subequations}
This all-pairs formulation introduces
\(N_\mathrm S K(K-1)\)
complex repetition equalities. For \(K\geq2\), this is \(K/2\) times as
many constraints as the single-reference formulation
in~\eqref{eq:coherency_repetition_constraints}.
Together with the per-mode binary-programmability constraints in~\eqref{eq:coherency_per_mode_binary}, these all-pairs constraints in~\eqref{eq:coherency_repetition_constraints_all_pairs} enforce one common load state among all modes with \(|x_{i,k}|+|u_{i,k}|>0\). Modes for which
\(x_{i,k}=u_{i,k}=0\)
impose no restriction because the state of that element is immaterial to their response. 

\subsection{Optimization Problems and Lifted SDR}
The coherency-synthesis problem requires the optimization of
\(\Cout(\mathbf x)\) over auxiliary vectors corresponding to one common
physical load vector. Denoting this exact physical feasible set by \(\mathcal X_{\mathrm{phys}}\), and assuming \(\Sout\triangleq\operatorname{tr}(\Cout)>0\) for all problems involving \(F\), we consider the four problems
\begin{subequations}
\label{eq:coherency_four_problems}
\begin{align}
    &\max_{\mathbf x\in\mathcal X_{\mathrm{phys}}}\quad U(\mathbf x),
    \label{eq:coherency_useful_only_problem}\\
    &\max_{\mathbf x\in\mathcal X_{\mathrm{phys}}}\quad U(\mathbf x)
    \quad\mathrm{s.t.}\quad F(\mathbf x)\geq F_{\min},
    \label{eq:coherency_useful_at_fidelity_problem}\\
    &\max_{\mathbf x\in\mathcal X_{\mathrm{phys}}}\quad F(\mathbf x)
    \quad\mathrm{s.t.}\quad U(\mathbf x)\geq U_{\min},
    \label{eq:coherency_fidelity_at_useful_problem}\\
    &\max_{\mathbf x\in\mathcal X_{\mathrm{phys}}}\quad F(\mathbf x).
    \label{eq:coherency_fidelity_only_problem}
\end{align}
\end{subequations}
For the bounds derived below, we represent this physical feasibility condition
using the per-mode binary-programmability constraints together with either of
the two repetition-constraint formulations introduced above.
The first and fourth problems in~\eqref{eq:coherency_four_problems} give the useful-strength-only and fidelity-only
limits, respectively, while the two constrained problems provide
complementary parameterizations of the useful-strength--fidelity Pareto
frontier.
Such complementary threshold sweeps expose tradeoffs that separate
single-objective limits cannot reveal; other Pareto-bound analyses were presented for characteristics of static antennas in~\cite{gustafsson2019tradeoff} and for output-wavefront synthesis with a coherently excited reconfigurable wave system in~\cite{del2026prototype}.

Whereas the concentration problem treated in Sec.~\ref{sec:concentration_bound} admitted a QCQP formulation in its auxiliary variable $\mathbf{y}$, the four problems in \eqref{eq:coherency_four_problems} do not generally admit direct QCQP
formulations in the auxiliary variable \(\mathbf x\). Specifically, although the prototype-feasibility
constraints and \(\Cout(\mathbf x)\) are quadratic in \(\mathbf x\), the
matrix square roots in \eqref{eq:coherency_fidelity} and
\eqref{eq:coherency_useful_strength} are not. We therefore first lift the
quadratic dependence by defining
\begin{equation}
    \mathbf Z\triangleq\mathbf x\mathbf x^\dagger
    \in\mathbb C^{N_\mathrm S K\times N_\mathrm S K}
\end{equation}
and partition \(\mathbf Z\) into \(N_\mathrm S\times N_\mathrm S\) blocks
\(\mathbf Z_{k\ell}\), where \(k,\ell=1,\ldots,K\). Let \(\mathcal Q\)
collect the binary-programmability and selected repetition constraints. We
write each \(q\in\mathcal Q\) as
\(q(\mathbf x)=\mathbf x^\dagger\mathbf R_q\mathbf x+
\mathbf x^\dagger\mathbf p_q+\mathbf s_q^\dagger\mathbf x+t_q=0\), where
\(\mathbf R_q\in\mathbb C^{N_\mathrm S K\times N_\mathrm S K}\),
\(\mathbf p_q,\mathbf s_q\in\mathbb C^{N_\mathrm S K}\), and
\(t_q\in\mathbb C\). Its affine lifted form is
\begin{equation}
    \mathcal L_q(\mathbf Z,\mathbf x)
    \triangleq
    \operatorname{tr}(\mathbf R_q\mathbf Z)
    +\mathbf x^\dagger\mathbf p_q
    +\mathbf s_q^\dagger\mathbf x+t_q=0.
    \label{eq:coherency_lifted_constraint}
\end{equation}
As in Sec.~\ref{sec:concentration_bound}, we impose separately the real and
imaginary parts of generally complex-valued equalities.

Recalling \(\mathbf X=\operatorname{unvec}(\mathbf x)\), based on~\eqref{eq:coherency_mode_factorization} we likewise express
the output coherency matrix affinely in the lifted variables as
\begin{equation}
    \mathcal C(\mathbf Z,\mathbf x)
    \triangleq{}
    \mathbf H_\mathrm e\mathbf H_\mathrm e^\dagger
    +\mathbf A\mathbf X\mathbf H_\mathrm e^\dagger
    +\mathbf H_\mathrm e\mathbf X^\dagger\mathbf A^\dagger
    +\mathbf A\left(\sum_{k=1}^{K}\mathbf Z_{kk}\right)\mathbf A^\dagger .
\label{eq:coherency_lifted_output}
\end{equation}
For \(\mathbf Z=\mathbf x\mathbf x^\dagger\),
\(\mathcal C(\mathbf Z,\mathbf x)=\Cout\) exactly. We obtain an SDR by
replacing this non-convex rank-one equality with its Schur-complement
relaxation
\begin{equation}
    \begin{bmatrix}
        \mathbf Z & \mathbf x\\
        \mathbf x^\dagger & 1
    \end{bmatrix}\succeq\mathbf0 .
    \label{eq:coherency_lifted_psd}
\end{equation}
We denote by \(\mathcal F_\mathrm{SDR}\) the resulting convex set of
\((\mathbf Z,\mathbf x,\Cout)\) satisfying
\(\mathcal L_q(\mathbf Z,\mathbf x)=0\) for every \(q\in\mathcal Q\),
\(\Cout=\mathcal C(\mathbf Z,\mathbf x)\), \(\Cout\succeq\mathbf0\), and
\eqref{eq:coherency_lifted_psd}.

It remains to represent the fidelity-based metrics in terms of
\((\mathbf Z,\mathbf x)\). To that end, we define the root fidelity for
positive-semidefinite matrices \(\mathbf C\) and \(\mathbf D\) as
\begin{equation}
    \mathfrak f(\mathbf C,\mathbf D)
    \triangleq
    \operatorname{tr}\sqrt{\sqrt{\mathbf D}\mathbf C\sqrt{\mathbf D}},
    \label{eq:coherency_root_fidelity}
\end{equation}
which admits the SDP representation~\cite[Sec.~2.1]{watrous2013simpler}
\begin{equation}
\begin{aligned}
    \mathfrak f(\mathbf C,\mathbf D)
    =\max_{\mathbf W}\quad&
    \operatorname{Re}\{\operatorname{tr}(\mathbf W)\}\\
    \mathrm{s.t.}\quad&
    \begin{bmatrix}
        \mathbf C & \mathbf W\\
        \mathbf W^\dagger & \mathbf D
    \end{bmatrix}\succeq\mathbf0 .
\end{aligned}
\label{eq:coherency_root_fidelity_sdp}
\end{equation}
Our two metrics thus become
\(U=\mathfrak f^2(\Cout,\rhostar)\) and
\(F=\mathfrak f^2(\Cout,\rhostar)/\Sout\).

\subsection{Complementary Pareto-Frontier Bounds}
We first derive a bound for
\eqref{eq:coherency_useful_at_fidelity_problem}. For \(\Sout>0\),
homogeneity gives
\(\mathfrak f(\Cout,\Sout\rhostar)=\Sout\sqrt{F}\); hence,
\(F\geq F_{\min}\) holds if and only if there exists a matrix
\(\mathbf W_F\) satisfying the corresponding semidefinite and trace
constraints in~\eqref{eq:coherency_useful_sdr}. We therefore solve
\begin{equation}
\begin{aligned}
    g_{U,\mathrm{SDR}}(F_{\min})
    =\max_{\substack{\mathbf Z,\mathbf x,\Cout,\\
                     \mathbf W_U,\mathbf W_F}}\quad&
    \operatorname{Re}\{\operatorname{tr}(\mathbf W_U)\}\\
    \mathrm{s.t.}\quad&
    (\mathbf Z,\mathbf x,\Cout)\in\mathcal F_\mathrm{SDR},\\
    &\begin{bmatrix}
        \Cout & \mathbf W_U\\
        \mathbf W_U^\dagger & \rhostar
      \end{bmatrix}\succeq\mathbf0,\\
    &\begin{bmatrix}
        \Cout & \mathbf W_F\\
        \mathbf W_F^\dagger & \Sout\rhostar
      \end{bmatrix}\succeq\mathbf0,\\
    &\operatorname{Re}\{\operatorname{tr}(\mathbf W_F)\}
      \geq\sqrt{F_{\min}}\,\Sout .
\end{aligned}
\label{eq:coherency_useful_sdr}
\end{equation}
Because maximizing the nonnegative root fidelity is equivalent to maximizing
its square, $U_\mathrm{SDR}(F_{\min})     \triangleq   g_{U,\mathrm{SDR}}^2(F_{\min})$
upper-bounds the useful strength achievable with \(F\geq F_{\min}\).
Consequently, \(U_\mathrm{SDR}(0)\) is the useful-strength-only bound for~\eqref{eq:coherency_useful_only_problem},
while sweeping \(F_{\min}\) gives one parameterization of the relaxed
Pareto frontier for~\eqref{eq:coherency_useful_at_fidelity_problem}.

We next derive a bound for the complementary problem
\eqref{eq:coherency_fidelity_at_useful_problem}, which maximizes \(F\)
subject to \(U\geq U_{\min}\). The normalization by \(\Sout\) makes this
a fractional problem. We remove this normalization with the
Charnes--Cooper transformation
~\cite{charnes1962programming,boyd2004convex}, following the same strategy
as in previous work on MIMO-operator synthesis for coherently excited
reconfigurable wave systems~\cite{del2026electromagnetic}.

Before applying the Charnes--Cooper transformation, we compute the auxiliary
total-output-power bound
\begin{equation}
\begin{aligned}
    \bar S
    \triangleq
    \max_{\mathbf Z,\mathbf x,\Cout}\quad
    & \operatorname{tr}(\Cout)\\
    \mathrm{s.t.}\quad
    & (\mathbf Z,\mathbf x,\Cout)
      \in\mathcal F_\mathrm{SDR}^{(\mathrm{ref})}.
\end{aligned}
\label{eq:strength_upper_bound}
\end{equation}
where \(\mathcal F_\mathrm{SDR}^{(\mathrm{ref})}\) denotes the lifted SDR feasible
set constructed with the single-reference repetition constraints.
The problem in~\eqref{eq:strength_upper_bound} is a convex SDP because
\(\Cout\) depends affinely on the lifted variables through
\(\mathcal C(\mathbf Z,\mathbf x)\), and the objective
\(\operatorname{tr}(\Cout)\) is linear. It needs to be solved only once per
scenario, before the Pareto sweep.
Since the all-pairs formulation includes the single-reference constraints,
its feasible set is a subset of
\(\mathcal F_\mathrm{SDR}^{(\mathrm{ref})}\). Consequently,
\(\bar S\) upper-bounds \(\Sout\) for both repetition formulations.

For \(\Sout>0\), we define
\begin{equation}
    \eta\triangleq\frac{1}{\Sout},\qquad
    \widetilde{\mathbf Z}\triangleq\eta\mathbf Z,\qquad
    \widetilde{\mathbf x}\triangleq\eta\mathbf x,\qquad
    \boldsymbol\rho\triangleq\eta\Cout .
    \label{eq:coherency_cc_variables}
\end{equation}
Then \(\operatorname{tr}(\boldsymbol\rho)=1\) and
\(\eta\geq\eta_\mathrm{min}\triangleq1/\bar S>0\). Defining
\(\widetilde{\mathbf X}\triangleq
\operatorname{unvec}(\widetilde{\mathbf x})\), the homogenized affine maps
are
\begin{equation}
\begin{aligned}
    \widetilde{\mathcal L}_q
    \triangleq{}&
    \operatorname{tr}(\mathbf R_q\widetilde{\mathbf Z})
    +\widetilde{\mathbf x}^\dagger\mathbf p_q
    +\mathbf s_q^\dagger\widetilde{\mathbf x}+t_q\eta,\\
    \widetilde{\mathcal C}
    \triangleq{}&
    \eta\mathbf H_\mathrm e\mathbf H_\mathrm e^\dagger
    +\mathbf A\widetilde{\mathbf X}\mathbf H_\mathrm e^\dagger
    +\mathbf H_\mathrm e\widetilde{\mathbf X}^\dagger\mathbf A^\dagger
    +\mathbf A\left(
        \sum_{k=1}^{K}\widetilde{\mathbf Z}_{kk}
      \right)\mathbf A^\dagger .
\end{aligned}
\label{eq:coherency_cc_affine_maps}
\end{equation}
Applying this transformation to the SDR of
\eqref{eq:coherency_fidelity_at_useful_problem} yields the following
single SDP:
\begin{equation}
\begin{aligned}
    g_{F,\mathrm{SDR}}(U_{\min})
    =\max_{\substack{\widetilde{\mathbf Z},
                     \widetilde{\mathbf x},\eta,\boldsymbol\rho,\\
                     \mathbf W_F,\mathbf W_U}}\quad&
    \operatorname{Re}\{\operatorname{tr}(\mathbf W_F)\}\\
    \mathrm{s.t.}\quad&
    \widetilde{\mathcal L}_q=0,\quad q\in\mathcal Q,\\
    &\boldsymbol\rho=\widetilde{\mathcal C},\quad
      \boldsymbol\rho\succeq\mathbf0,\quad
      \operatorname{tr}(\boldsymbol\rho)=1,\\
    &\eta\geq\eta_\mathrm{min},\quad
      \begin{bmatrix}
        \widetilde{\mathbf Z} & \widetilde{\mathbf x}\\
        \widetilde{\mathbf x}^\dagger & \eta
      \end{bmatrix}\succeq\mathbf0,\\
    &\begin{bmatrix}
        \boldsymbol\rho & \mathbf W_F\\
        \mathbf W_F^\dagger & \rhostar
      \end{bmatrix}\succeq\mathbf0,\\
    &\begin{bmatrix}
        \boldsymbol\rho & \mathbf W_U\\
        \mathbf W_U^\dagger & \eta\rhostar
      \end{bmatrix}\succeq\mathbf0,\\
    &\operatorname{Re}\{\operatorname{tr}(\mathbf W_U)\}
      \geq\eta\sqrt{U_{\min}} .
\end{aligned}
\label{eq:coherency_fidelity_cc_sdr}
\end{equation}
By~\eqref{eq:coherency_root_fidelity_sdp}, the largest value of
\(\operatorname{Re}\{\operatorname{tr}(\mathbf W_U)\}\) permitted by the
block positive-semidefinite constraint involving \(\mathbf W_U\) is
\begin{equation}
    \mathfrak f(\boldsymbol\rho,\eta\rhostar)
    =
    \sqrt{\eta F}.
    \label{eq:coherency_cc_useful_amplitude}
\end{equation}
Hence, the constraint
\(\operatorname{Re}\{\operatorname{tr}(\mathbf W_U)\}
\geq\eta\sqrt{U_{\min}}\) is feasible if and only if
\(F/\eta=U\geq U_{\min}\). Consequently, $F_\mathrm{SDR}(U_{\min})     \triangleq     g_{F,\mathrm{SDR}}^2(U_{\min})$ upper-bounds the fidelity achievable in
\eqref{eq:coherency_fidelity_at_useful_problem}.
Setting \(U_{\min}=0\) yields the fidelity-only bound for
\eqref{eq:coherency_fidelity_only_problem}. The lower bound
\(\eta_\mathrm{min}=1/\bar S>0\) excludes the degenerate case
\(\eta=0\), which cannot correspond to any finite physical output power.

As an alternative to
\eqref{eq:coherency_fidelity_cc_sdr}, the monotonicity of
\(U_\mathrm{SDR}(F_{\min})\) permits the equivalent evaluation
\begin{equation}
    F_\mathrm{SDR}(U_{\min})
    =
    \sup\left\{
        F_{\min}\in[0,1]:
        U_\mathrm{SDR}(F_{\min})\geq U_{\min}
    \right\}
    \label{eq:coherency_fidelity_bisection}
\end{equation}
by bisection~\cite{boyd2004convex}. However, the direct Charnes--Cooper formulation
is computationally more efficient: after one reusable solve for
\(\bar S\), it requires solving only one SDP per useful-strength threshold,
whereas the bisection implementation requires multiple SDP solves per threshold.

\subsection{Validity and Ambiguity Insensitivity}

The quantities \(U_\mathrm{SDR}(0)\),
\(U_\mathrm{SDR}(F_{\min})\),
\(F_\mathrm{SDR}(U_{\min})\), and
\(F_\mathrm{SDR}(0)\) upper-bound the optimal objective values of
\eqref{eq:coherency_useful_only_problem},
\eqref{eq:coherency_useful_at_fidelity_problem},
\eqref{eq:coherency_fidelity_at_useful_problem}, and
\eqref{eq:coherency_fidelity_only_problem}, respectively.
Indeed, every admissible binary configuration yields a rank-one lifted
point that is feasible for the corresponding SDP. The single-reference
formulation may additionally relax the exact common-load constraint in
degenerate cases where the reference mode does not excite a given element,
whereas the all-pairs formulation avoids this reference-mode dependence.
Both formulations then apply the usual SDR rank relaxation, which permits
higher-rank moments. The Charnes--Cooper transformation is an exact
rescaling for \(\Sout>0\) and introduces no additional relaxation.

Within the reciprocal proxy-MNT representation used in this work, these four
bounds are also insensitive to the ambiguities of experimentally
estimated proxy-MNT parameters. Specifically, the invertible affine variable
transformations and congruence transformations established for the
repetition-constrained MIMO formulation in the ambiguity appendix
of~\cite{del2026electromagnetic} preserve the lifted feasibility constraints
and the physical \(\Cout\). They consequently preserve \(\Sout\),
\(\bar S\), \(\eta\), the normalized coherency matrix, and both block
positive-semidefinite fidelity constraints. Hence, the complete relaxed
Pareto frontier is independent of the chosen reciprocity-preserving
proxy-MNT representation.

\section{Discrete-Optimization Benchmarks}
\label{sec:discrete_optimization_benchmarks}

To assess how tight the SDR-based bounds are in practice, we compare them with
feasible binary configurations found by several discrete-optimization procedures. A
small gap between the bound and the best discrete configuration indicates that
the bound is nearly attainable; a larger gap can reflect either a loose
relaxation or the inability of the tested discrete optimizers to find the best
binary configuration. The discrete-optimization algorithms summarized in this section are well-known and only used as feasible benchmarks. In all cases, once a binary
load vector \(\mathbf r\) is chosen, we evaluate it with the proxy MNT model
in~\eqref{eq:mnt_map}; the score is \(P_1\) in the concentration study, whereas
for coherency synthesis CD and GA maximize
\(J_w=(1-w)U/U_{\mathrm{SDR}}(0)+wF\) using the same 25 prescribed weights
\(w\in[0,1]\), and we report the resulting \((F,U)\) pairs.

For sufficiently small \(N_\mathrm S\), we enumerate all
\(2^{N_\mathrm S}\) configurations and use the best one as the exhaustive-search
(ES) reference. This is the only discrete benchmark that certifies global
optimality over the binary feasibility set. The enumeration can be implemented in
Gray-code order so that consecutive configurations differ by one load only,
allowing the MNT inverse to be updated efficiently with the Woodbury
identity~\cite{prod2023efficient}.

In addition, for all values of \(N_\mathrm{S}\), we use coordinate ascent
(CD), a genetic algorithm (GA), and a projection of the SDR optimizer (P-SDR).
CD is a local optimizer that we run with multiple initializations to
reduce the sensitivity to local optima of the non-convex binary problem. Starting
from one binary configuration, CD tests single-load flips and accepts only
flips that improve the objective.\footnote{One CD sweep tests one possible flip of each of the \(N_\mathrm{S}\) binary load states, in random order. We allow at most 50
sweeps per initialization and use at least ten distinct initializations; we stop earlier when a complete sweep accepts no flip. We evaluate single-flip candidates with computationally efficient Woodbury updates~\cite{prod2023efficient}.} 
GA is a population-based stochastic optimizer that, starting from an initial
population of configurations, repeatedly selects high-performing
configurations, combines them by crossover, mutates individual load states,
and keeps elite configurations between generations. We polish the best GA
outcome with CD.\footnote{Our GA settings use a population size of at least 60, at most 80 generations, an elite count of 5, crossover probability 0.85, and mutation probability \(\max(1/N_\mathrm{S},0.01)\).}
Finally, P-SDR
turns the continuous SDR optimizer into a feasible binary configuration by
projecting its relaxed internal state onto the two admissible load
states.\footnote{The projection is based on the MNT fixed-point relation in
\eqref{eq:reciprocal_auxiliary_fixed_point} and its matrix-valued analogue
associated with~\eqref{eq:coherency_internal_matrix}. After solving
\eqref{eq:concentration_sdr} for concentration, or an unscaled
coherency-synthesis SDR such as~\eqref{eq:coherency_useful_sdr}, we take
the optimized auxiliary variables \(\mathbf y^\star\) and
\(\mathbf X^\star=\operatorname{unvec}(\mathbf x^\star)\), respectively.
For Charnes--Cooper solves of~\eqref{eq:coherency_fidelity_cc_sdr}, the
corresponding unscaled auxiliary variable is instead
\(\mathbf X^\star=\operatorname{unvec}(\widetilde{\mathbf x}^\star/\eta^\star)\).
These auxiliary variables need not correspond to a binary load vector after
relaxation. For each element \(i\), we compare the
fixed-point residuals obtained by substituting \(r_i=\alpha\) and
\(r_i=\beta\), and choose the load state with the smaller residual. Whenever the auxiliary variable has multiple columns, the squared residual magnitudes are summed over
those columns to enforce one shared physical load state.}

Besides P-SDR, we also consider two additional techniques to extract feasible candidate designs from the SDR optimizer. While P-SDR maps the optimized auxiliary variable returned by the SDR directly to binary loads, our dominant-eigenvector SDR projection (DE-SDR) first forms a candidate auxiliary variable from the dominant eigenvector of the augmented lifted optimizer and then applies the same binary projection.\footnote{We form the augmented lifted optimizer \(\mathbf M^\star=
\left[\begin{smallmatrix}
\mathbf Y^\star & \mathbf y^\star\\
(\mathbf y^\star)^\dagger & 1
\end{smallmatrix}\right]\)
for the concentration problem, with the analogous replacement
\((\mathbf Y^\star,\mathbf y^\star)\mapsto(\mathbf Z^\star,\mathbf x^\star)\)
for coherency synthesis. Then, we extract the dominant eigenvector
\(\mathbf u=[\mathbf u_1^\top\,\,u_0]^\top\). If \(|u_0|>10^{-12}\), we form
the candidate auxiliary variable \(\mathbf u_1/u_0\); otherwise, we use the scaled candidate \(\sqrt{\lambda_{\max}(\mathbf M^\star)}\mathbf u_1\). Finally, we apply the same
binary residual projection as in P-SDR.}
A complementary strategy is motivated by the idea that an SDR optimizer that is
closer to rank one should be easier to convert into a high-performing feasible
binary design~\cite{liu_rankOneSolutions2019,gertler2025many,chao2026blueprints}. Following~\cite{liu_rankOneSolutions2019}, our rank-penalized SDR projection
(RP-SDR) iteratively re-solves the SDR with an additional linearized penalty that approximates \(\mathrm{tr}(\mathbf M^\star)-\lambda_{\max}(\mathbf M^\star)\), where \(\mathbf M^\star\) denotes the augmented lifted optimizer of the current
rank-penalized SDR. After this
rank-promoting step, we apply the same dominant-eigenvector extraction and
binary residual projection as in DE-SDR.\footnote{In our implementation,
RP-SDR starts from the unpenalized SDR optimizer, uses an initial penalty
weight \(10^{-3}\), allows at most two rank-penalized SDR solves, and
increases the penalty weight by a factor of 10 between solves.}

\begin{figure*}
    \centering
    \includegraphics[width=2\columnwidth]{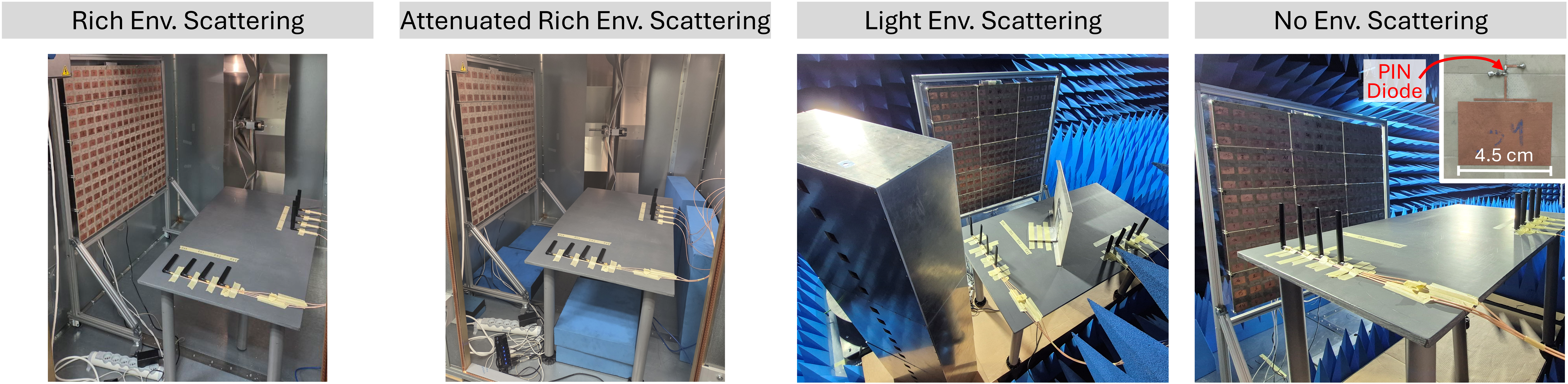}
    \caption{Photographic images of the four considered programmable MIMO wave systems, based on RIS-parametrized radio environments.}
    \label{Fig1}
\end{figure*}

\section{Experimental Results}
\label{sec_Results}

In this section, we evaluate our bounds on fabricated prototypes of programmable wave systems. We describe our experimental prototypes of reconfigurable wave systems in Sec.~\ref{subsec:experimental_setups}, our corresponding concentration bounds in Sec.~\ref{subsec:concentration_results}, and our corresponding synthesis bounds in Sec.~\ref{subsec:pareto_results}.

\subsection{Experimental Setups}
\label{subsec:experimental_setups}

We conduct all experiments at 2.45~GHz.
The main ingredients of our four prototype systems are an RIS, a transmit array, and a receive array. We deploy these in four different radio environments, yielding four distinct RIS-parametrized MIMO systems. Changing the propagation conditions while leaving the RIS hardware unchanged gives us access to different regimes of coupling between the RIS elements, because environmental scattering contributes substantially to the interactions between RIS elements~\cite{rabault2023tacit,del2025experimentalreducedrank}. Each antenna array consists of four Wi-Fi antennas, arranged in parallel with half-wavelength spacing. Our RIS consists of 225 elements of approximately half-wavelength size and operates around \(2.45~\mathrm{GHz}\). Each RIS element comprises an electrically small PIN diode so that the RIS element can be switched between two states; further details on the RIS design can be found in~\cite{KDF14,ahmed2025over}. The PIN diode's small electrical size is consistent with our lumped-load description of the tunable component in Sec.~\ref{sec:system_model}~\cite{largeRIS_TCOM,delHougne_frozenDifferential2026}. We use at most \(N_\mathrm S=100\) RIS elements in the present study and leave the other RIS elements in a fixed reference state. 

Our four propagation environments are displayed in Fig.~\ref{Fig1}. 
For our first setup, we use an unloaded reverberation chamber as the propagation environment, characterized by strong environmental scattering. For our second setup, we introduce absorbing material into the same chamber to attenuate the reverberation. For our third setup, we use an anechoic chamber together with several metallic scattering objects as the propagation environment. For our fourth setup, we remove these additional scattering objects from the anechoic chamber. During all measurements in the reverberation chamber, the mechanical mode stirrer seen in Fig.~\ref{Fig1} remains in the same fixed position. 

In each of the four radio environments, we use an eight-port vector network analyzer to measure the complete \(4\times4\) end-to-end channel matrix. As discussed earlier, direct experimental access to the MNT parameters of~\eqref{eq:mnt_map} is not available. In particular, the PIN diodes are soldered onto the RIS elements, so neither the corresponding ``virtual'' ports nor the corresponding tunable loads can be connected to the VNA. Moreover, with \(N_\mathrm S=100\), the static MNT subsystem contains \(N_\mathrm T+N_\mathrm R+N_\mathrm S=108\) ports, whereas our VNA only has eight measurement ports. Obtaining the MNT parameters of~\eqref{eq:mnt_map}  through full-wave simulation would also be impractical: the exact geometry and material composition of the radio environments are not known with sufficient detail, and the overall structures are electrically very large. Thus, we instead infer an equivalent proxy MNT description separately for each of the four environments from end-to-end measurements obtained for known RIS states, using the estimation procedure in~\cite{ContRIS_LWC}. The MNT parameters are typically not uniquely identifiable, such that the resulting proxy MNT parameters generally do not coincide with the ``true'' MNT parameters. The underlying ambiguities are discussed in the appendices
of~\cite{del2026electromagnetic,del2026prototype}. Nonetheless, the proxy MNT
parameters are operationally equivalent as long as they generate the same
end-to-end response for every allowed RIS configuration; within the
reciprocity-preserving proxy class used here, the corresponding ambiguities
leave the bounds considered here unchanged.

We assess the accuracy of each proxy MNT model based on its ability to predict the end-to-end channel matrix for unseen RIS configurations. For this purpose, we use the accuracy metric \(\zeta\) of~\cite{ContRIS_LWC}. Its interpretation is analogous to an SNR: the measured channel response plays the role of the signal, whereas the difference between the measured response and the proxy-model prediction plays the role of noise. Thus, higher \(\zeta\) indicates a smaller modeling error. For our four setups, we obtain \(47.4\), \(55.3\), \(39.1\), and \(42.7~\mathrm{dB}\), respectively. These values correspond to relative prediction errors of roughly one percent or less. This level of accuracy is noteworthy given the dimensionality of the fitted proxy MNT model; already the reciprocal RIS coupling matrix contains
\(N_\mathrm S(N_\mathrm S+1)/2=5050\) independent entries for \(N_\mathrm S=100\). Since we evaluate our bounds based on the calibrated proxy MNT model, model-estimation uncertainty is not propagated into our bounds; however, the above prediction errors indicate only small residual model mismatch.

Our proxy MNT parameter estimation procedure initially yields models with four transmitting and four receiving antennas, as well as 100 RIS elements. 
When we wish to consider only a subset of the transmitting and/or receiving antennas, we terminate the unused antennas in matched loads. Since a matched load has zero reflection coefficient, incorporating these terminations into the calibrated model only requires selecting the appropriate subblocks. When we wish to consider only a subset of the RIS elements, we fix the unused RIS elements to their reference state. Because within our proxy MNT parameters that reference state is fixed to a matched load (see details in~\cite{ContRIS_LWC}), incorporating these fixed states into the calibrated model also only requires selecting the appropriate subblocks.

\subsection{Concentration Results}
\label{subsec:concentration_results}

\begin{figure*}[t]
    \centering
    \includegraphics[width=2\columnwidth]
    {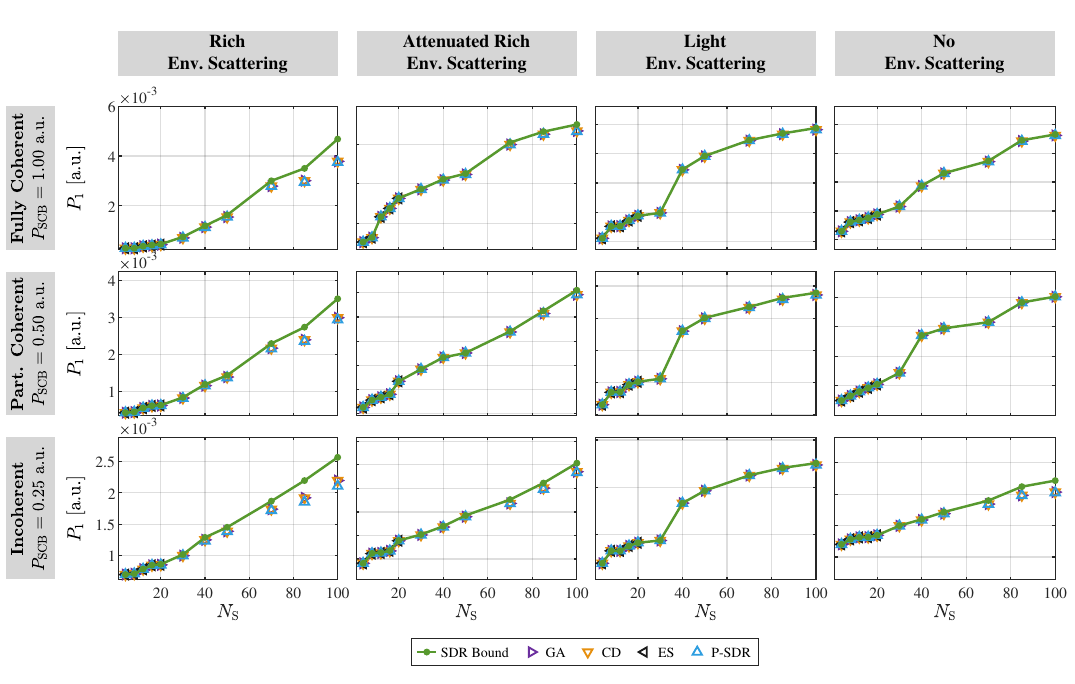}
    \caption{Prototype-aware upper bound and discrete optimization
    outcomes for concentrating a four-port input into
    output port~1. Columns correspond to the four experimental radio
    environments shown in Fig.~\ref{Fig1}, and rows correspond to the fully coherent (FC),
    partially coherent (PC), and incoherent (IC) input cases defined in
    \eqref{eq:input_coherency_scenarios}. We perform exhaustive search (ES) only for
    \(N_\mathrm{S}\leq20\).}
    \label{Fig2}
\end{figure*}

We evaluate our concentration bound from
Sec.~\ref{sec:concentration_bound} using \(N_\mathrm{T}=4\) input ports
and \(N_\mathrm{R}=1\) output port, choosing output port~1 in every case. We vary the
number of programmable RIS elements over
\(N_\mathrm{S}\in\{4,8,12,16,20,30,40,50,70,85,100\}\).

We consider three representative input coherency
matrices whose trace we normalize such that \(\mathrm{tr}(\Cin)=1\).
This normalization fixes
the incident-power scale; accordingly, we report \(P_1\) and
\(P_{\mathrm{SCB}}\) in arbitrary units.
Specifically, defining
\begin{equation}
\begin{aligned}
\mathbf u
&=\frac{1}{2}
\begin{bmatrix}
1 & e^{\jmath 0.35\pi} & e^{-\jmath 0.20\pi}
  & e^{\jmath 0.65\pi}
\end{bmatrix}^{\mathsf T},\\
\mathbf u_{13}
&=\frac{1}{\sqrt{2}}
\begin{bmatrix}
1 & 0 & e^{\jmath 0.55\pi} & 0
\end{bmatrix}^{\mathsf T},\\
\mathbf u_{24}
&=\frac{1}{\sqrt{2}}
\begin{bmatrix}
0 & 1 & 0 & e^{-\jmath 0.20\pi}
\end{bmatrix}^{\mathsf T},
\end{aligned}
\label{eq:input_coherency_modes}
\end{equation}
we construct
\begin{equation}
\begin{aligned}
\Cin^{(\mathrm{FC})}
&=\mathbf u\mathbf u^\dagger,\\
\Cin^{(\mathrm{PC})}
&=0.30\,\mathbf u\mathbf u^\dagger
+0.70\left(
0.55\,\mathbf u_{13}\mathbf u_{13}^\dagger
+0.45\,\mathbf u_{24}\mathbf u_{24}^\dagger
\right),\\
\Cin^{(\mathrm{IC})}
&=\frac{1}{4}\mathbf I_4 .
\end{aligned}
\label{eq:input_coherency_scenarios}
\end{equation}
The corresponding eigenvalue spectra are
\((1,0,0,0)\), \((0.50,0.36,0.14,0)\), and
\((0.25,0.25,0.25,0.25)\), and the ranks are \(1\), \(3\), and \(4\),
respectively. The fully coherent (FC) case models a synchronized excitation occupying one coherent mode across all four ports. The partially coherent (PC) case combines a global coherent mode with two statistically independent port-pair modes and represents, for example, several source groups that retain coherence internally but are not mutually synchronized. The incoherent (IC) case assigns equal power to four
uncorrelated channels and represents the limiting case of spatially
incoherent excitation. 
These scenarios therefore differ not only in coherency rank, but also
in eigenvalue spectrum and coherent-mode orientation and, for the PC
case, in the average powers incident through the individual input ports.

We display in Fig.~\ref{Fig2} our prototype-aware SDR bound on the
achievable concentrated power and the feasible outcomes obtained with the
discrete optimizations of
Sec.~\ref{sec:discrete_optimization_benchmarks}. At
\(N_\mathrm S=100\), the prototype-aware bounds amount to only
\(0.01\%\)--\(1.03\%\) of the architecture-independent
\(P_{\mathrm{SCB}}\) from~\eqref{eq:scattering_concentration_bound}, reflecting the substantial restrictions imposed by
the experimental prototypes. We also evaluated the transmission-scaled SCB from~\eqref{eq:transmission_scaled_scb}. 
For IC, it coincides with the bound obtained via the direct concentration SDR
in~\eqref{eq:concentration_sdr}, because
\(\Cin^{(\mathrm{IC})}=\mathbf I_4/4\) and
\(\bar\eta_{\mathrm c}<1\) in all examined cases, making both bounds equal to
\(\bar\eta_{\mathrm c}/4\). For FC and PC, the transmission-scaled SCB retains
only \(\lambda_{\max}(\Cin)\), whereas the direct SDR exploits the full input
coherency matrix and is tighter in all examined cases. At \(N_\mathrm S=100\),
the scaled-to-direct bound ratio ranges from \(2.19\) to \(9.51\) for FC and
from \(1.47\) to \(2.87\) for PC. This gap quantifies the loss of modal-alignment
information incurred by replacing the full \(\Cin\) with
\(\lambda_{\max}(\Cin)\mathbf I\). We therefore omit the transmission-scaled
SCB from Fig.~\ref{Fig2}.

For \(N_\mathrm{S}\leq20\), exhaustive search determines the global
binary optimum and directly reveals the SDR relaxation gap. The largest
ratio between the SDR bound and the exhaustive-search result is \(1.048\)
in rich environmental scattering and at most \(1.0017\) in the other
three environments. Among the scalable discrete optimizations, CD, GA, and P-SDR yield essentially equivalent results in most cases. Specifically, CD and GA
differ by less than \(0.1\%\) at all 132 plotted points, while all three
methods remain within \(1\%\) of the best discrete result at 120 of
these points. P-SDR is occasionally slightly inferior in rich
environmental scattering; its largest deficit is \(4.0\%\), observed
for the IC input at \(N_\mathrm S=100\).

For larger \(N_\mathrm{S}\), exhaustive search is computationally
infeasible. We therefore interpret the difference between the SDR bound
and the best discrete result as a \emph{certification gap}, which may
contain both SDR-relaxation slack and residual suboptimality of the
discrete algorithms. Despite this conservative interpretation, the observed certification
gaps demonstrate the remarkable tightness of our prototype-aware bound.
Across all 132 plotted points, the gap does not exceed \(1\%\) in 101
cases and \(3\%\) in 115 cases. Even in the least favorable case, the
best discrete configuration attains \(80.89\%\) of the SDR bound,
corresponding to a bound-to-achieved-power ratio of only \(1.24\).
Thus, the bound remains practically informative throughout all
considered input coherencies, radio environments, and values of
\(N_\mathrm S\). At \(N_\mathrm{S}=100\), this gap equals
\(19.11\%\), \(14.59\%\), and \(14.39\%\) in rich environmental
scattering for the FC, PC, and IC inputs, respectively. The gaps become
much smaller in the attenuated-rich and light-scattering environments,
where they do not exceed \(2.94\%\). Environmental scattering richness
thus strongly influences the tightness of the bound.

The gap does not, however, depend on the radio environment alone. In
the no-environmental-scattering case, the gap is \(0.29\%\) for the FC
input and \(0.06\%\) for the PC input, but \(2.23\%\) for the IC input.
The near-vanishing FC gap is consistent with previous observations that
SDR bounds are often especially tight in free-space or
no-environmental-scattering settings for related objectives, including
SISO channel-gain maximization, MIMO aggregate-gain and
operator-synthesis objectives, and wavefront synthesis
~\cite{salmi2026electromagnetically,del2026electromagnetic,del2026prototype}.
The IC result nevertheless shows that removing environmental scattering
does not by itself guarantee a near-vanishing gap for an arbitrary input
coherency matrix. More generally, Fig.~\ref{Fig2} reveals no monotonic
relation between the gap and the rank of \(\Cin\). The rank alone is
therefore insufficient to predict tightness; the full input coherency
matrix and its relation to the programmable transmission response
matter.

Finally, we also evaluated the DE-SDR and RP-SDR extraction schemes for optimized configurations described in Sec.~\ref{sec:discrete_optimization_benchmarks}. DE-SDR reproduced P-SDR within numerical
precision, whereas RP-SDR was \(0.0054\%\) lower in one case. Subsequent CD or GA polishing of these
configurations never improved upon the best result obtained with
standard CD or GA. For this reason, we omitted the corresponding markers from Fig.~\ref{Fig2} for visual clarity.

\subsection{Synthesis Pareto-Frontier Results}
\label{subsec:pareto_results}

\begin{figure*}[t]
    \centering
    \includegraphics[width=2\columnwidth]{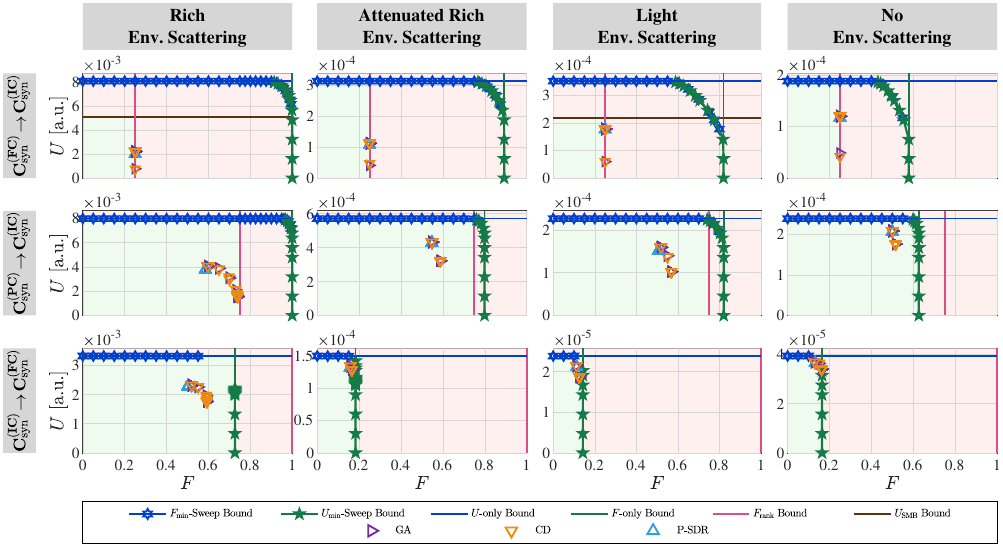}
    \caption{Upper bounds and discrete optimization outcomes
    for synthesizing selected \(4\times4\) target coherency matrices. Columns
    correspond to the four experimental radio environments shown in
    Fig.~\ref{Fig1}, and rows correspond to the transformations defined in
    \eqref{eq:synthesis_coherency_scenarios}. Blue and dark green curves result
    from the fidelity- and useful-strength-threshold sweeps in
    \eqref{eq:coherency_useful_sdr} and
    \eqref{eq:coherency_fidelity_cc_sdr}, respectively.
    Pink vertical lines show the architecture-independent rank-based fidelity
    bounds in \eqref{eq:rank_fidelity_bound}. Brown horizontal lines show the
    transmission-scaled useful-strength bounds in
    \eqref{eq:transmission_scaled_maximum_brightness}. Light red regions are
    certified infeasible, whereas light green regions are not excluded by the
    evaluated bounds.}
    \label{Fig3}
\end{figure*}

We evaluate our bounds on coherency synthesis for \(N_\mathrm{T}=N_\mathrm{R}=4\) and
\(N_\mathrm{S}=100\). With
\begin{equation}
\begin{aligned}
\mathbf v_\mathrm{F}
&=\frac{1}{2}
\begin{bmatrix}
1 & e^{\jmath 0.85\pi/3} & e^{\jmath 1.70\pi/3} & e^{\jmath 0.85\pi}
\end{bmatrix}^{\mathsf T},\\
\mathbf v_\mathrm{a}
&=\frac{1}{2}
\begin{bmatrix}
1 & e^{\jmath 0.35\pi} & e^{-\jmath 0.20\pi} & e^{\jmath 0.65\pi}
\end{bmatrix}^{\mathsf T},\\
\mathbf v_{12}
&=\frac{1}{\sqrt{2}}
\begin{bmatrix}1 & e^{\jmath 0.30\pi} & 0 & 0\end{bmatrix}^{\mathsf T},\\
\mathbf v_{34}
&=\frac{1}{\sqrt{2}}
\begin{bmatrix}0 & 0 & 1 & e^{-\jmath 0.45\pi}\end{bmatrix}^{\mathsf T},
\end{aligned}
\label{eq:synthesis_coherency_modes}
\end{equation}
we define the three unit-trace coherency matrices
\begin{equation}
\begin{aligned}
\mathbf C_\mathrm{syn}^{(\mathrm{FC})}
&=\mathbf v_\mathrm{F}\mathbf v_\mathrm{F}^\dagger,\\
\mathbf C_\mathrm{syn}^{(\mathrm{PC})}
&=0.25\,\mathbf v_\mathrm{a}\mathbf v_\mathrm{a}^\dagger
+0.75\bigl(0.58\,\mathbf v_{12}\mathbf v_{12}^\dagger
+0.42\,\mathbf v_{34}\mathbf v_{34}^\dagger\bigr),\\
\mathbf C_\mathrm{syn}^{(\mathrm{IC})}
&=\frac{1}{4}\mathbf I_4.
\end{aligned}
\label{eq:synthesis_coherency_scenarios}
\end{equation}
Their eigenvalue
spectra are, respectively, \((1,0,0,0)\),
\((0.602,0.329,0.069,0)\), and \((0.25,0.25,0.25,0.25)\). Our various bounds on the representative transformations \(\mathrm{FC}\!\rightarrow\!\mathrm{IC}\),
\(\mathrm{PC}\!\rightarrow\!\mathrm{IC}\), and
\(\mathrm{IC}\!\rightarrow\!\mathrm{FC}\) are displayed in  Fig.~\ref{Fig3}. 
For non-fully-coherent inputs, we use the single-reference formulation for the SDR-based bounds in Fig.~\ref{Fig3} because of its lower computational cost. Matched spot checks with the all-pairs formulation produced only limited tightening, reaching $6.4\%$ in the most pronounced case.

\begin{table}[t]
\centering
\footnotesize
\caption{Architecture-independent and prototype-aware fidelity-only bounds.
The prototype-aware entries are ordered as rich, attenuated-rich, light, and
no environmental scattering.}
\label{tab:synthesis_fidelity_bounds}
\setlength{\tabcolsep}{3.5pt}
\begin{tabular}{c c c}
\hline
Transformation & \(F_\mathrm{rank}\) & \(F_\mathrm{SDR}\)\\
\hline
FC\(\rightarrow\)IC & 0.25 & 1.00 / 0.89 / 0.82 / 0.58\\
PC\(\rightarrow\)IC & 0.75 & 1.00 / 0.80 / 0.82 / 0.63\\
IC\(\rightarrow\)FC & 1.00 & 0.73 / 0.18 / 0.14 / 0.16\\
\hline
\end{tabular}
\end{table}

Interestingly, in some cases the architecture-independent fidelity bound from~\eqref{eq:rank_fidelity_bound} remains
quantitatively informative despite ignoring the prototype. Specifically,
it is nontrivial when
\(\operatorname{rank}(\rhostar)>
\min\{\operatorname{rank}(\Cin),N_\mathrm{R}\}\), i.e., when the target
requires more mutually incoherent modes than a linear transformation of the
input can produce. For
\(\mathrm{FC}\!\rightarrow\!\mathrm{IC}\), the fact that the
rank cannot increase under a linear transformation limits the fidelity to
\(F_\mathrm{rank}=0.25\), which is tighter than all four prototype-aware \(F\)-only bounds
summarized in Table~\ref{tab:synthesis_fidelity_bounds}. For
\(\mathrm{PC}\!\rightarrow\!\mathrm{IC}\), the corresponding rank-only
ceiling of \(F_\mathrm{rank}=0.75\) is tighter in the first three environments, whereas the
prototype-aware value \(0.63\) is tighter without environmental scattering.
Conversely, $F_\mathrm{rank}$ trivially equals unity for
\(\mathrm{IC}\!\rightarrow\!\mathrm{FC}\), for which the prototype-aware
bounds are tighter in all four environments. 

The unscaled architecture-independent useful-strength bounds from~\eqref{eq:maximum_brightness_useful_strength} equal
\(0.25\), \(0.65\), and \(0.25\) for the three transformations and remain
loose for our rather lossy prototype systems. The SDR-certified transmission factors
\(\bar\eta_{\mathrm{tr}}\) range from
\(8.73\times10^{-4}\) to \(2.05\times10^{-2}\) and allow us to substantially tighten
these architecture-independent bounds based on~\eqref{eq:transmission_scaled_maximum_brightness}. The resulting
\(U_{\mathrm{SMB}}\) improves upon the direct
prototype-aware \(U\)-only SDR in two of the twelve cases considered in Fig.~\ref{Fig3}. For
\(\mathrm{FC}\!\rightarrow\!\mathrm{IC}\), it lowers the ceiling from
\(8.10\times10^{-3}\) to \(5.12\times10^{-3}\) in rich environmental
scattering and from \(3.52\times10^{-4}\) to \(2.18\times10^{-4}\) in the
light-scattering environment, reductions of \(36.8\%\) and \(37.9\%\),
respectively. 
The direct prototype-aware SDR remains tighter in the other ten cases.

The feasible outcomes confirm that some architecture-independent bounds are attainable despite the prototype constraints. In particular, all four
environments attain \(F=0.25\) for
\(\mathrm{FC}\!\rightarrow\!\mathrm{IC}\), thereby saturating the rank-only
ceiling. For \(\mathrm{PC}\!\rightarrow\!\mathrm{IC}\), the largest achieved
fidelities are \(0.739\), \(0.588\), \(0.568\), and \(0.516\) across the four
environments; the rich
environment therefore approaches the rank-only ceiling particularly closely.
For \(\mathrm{IC}\!\rightarrow\!\mathrm{FC}\), the corresponding values are
\(0.592\), \(0.165\), \(0.126\), and \(0.163\). The last value nearly attains
the prototype-aware \(F\)-only bound of \(0.165\), whereas the rich environment
retains a more visible certification gap relative to its bound of \(0.726\).
The vertical scales also
show that the useful-strength bounds depend strongly on the environment: the
\(U\)-only values range from \(2.38\times10^{-5}\) to
\(8.10\times10^{-3}\) across
the displayed panels.

The different feasible optimizers yield very similar frontiers. CD and GA
produce identical \((F,U)\) pairs for 288 of the 300 matched scalarizations in
Fig.~\ref{Fig3}; over the remaining pairs, their largest absolute differences
are \(9.9\times10^{-4}\) in fidelity and \(8.0\times10^{-5}\) in useful
strength.
P-SDR also tracks the realized frontier closely, although its useful strength
can be up to \(8.8\%\) below the best discrete result among the common feasible
points of the fidelity-threshold sweep.

Beyond the agreement between discrete optimizers, Fig.~\ref{Fig3} also
shows that the bounds themselves are often rather tight. This is especially
clear for the \(\mathbf C_\mathrm{syn}^{(\mathrm{FC})}\to
\mathbf C_\mathrm{syn}^{(\mathrm{IC})}\)
transformation: because a coherent input yields an output of rank at most one
under any deterministic linear transformation, the architecture-independent fidelity
ceiling is \(F\leq 1/4\), and the optimized configurations essentially attain
this ceiling in all four environments. The shape of the certified Pareto
frontier nevertheless varies noticeably between panels. Some frontiers are
nearly rectangular, indicating that high fidelity can be maintained over a
broad range of useful-strength levels, whereas others are visibly curved,
indicating a stronger tradeoff between useful strength and coherency-structure
fidelity.

Altogether, the fully prototype-aware SDR is not automatically the tightest bound in every direction of the Pareto plane: in some cases, the architecture-independent rank/fidelity bound or the transmission-scaled maximum-brightness bound is more restrictive. This indicates that the SDR relaxation does not retain all spectral constraints implied by deterministic linear scattering. These observations highlight the complementarity of the different bounds and motivate defining a certified outer bound on the feasible region as the intersection of all evaluated bounds.

\section{Conclusion}
\label{sec:Conclusion}

To summarize, we have established certified limits on how concrete real-world reconfigurable wave systems can manipulate the second-order statistics of partially coherent microwaves. Our experimental results show that prototype constraints can reduce achievable performance far below architecture-independent limits, while also revealing that no single bounding strategy is uniformly strongest: spectral, scalar-refined, and fully prototype-aware bounds provide complementary limits. The generally small gaps between the tightest bound and the best optimized feasible performance indicate that these bounds can meaningfully distinguish algorithmic limitations from genuine hardware limitations. More broadly, this work provides a framework for assessing and designing reconfigurable wave systems for wave-domain processing and RF energy harvesting under partially coherent or incoherent excitation.

\section*{Acknowledgment}

P.d.H. acknowledges stimulating discussions with I.~Liberal and C.~Roques-Carmes. 
P.d.H. acknowledges I.~Ahmed, F.~Boutet, and C.~Guitton who, under P.d.H.'s supervision, previously built the RIS prototype used in part of the experimental study. P.d.H. also acknowledges J.~Sol for technical support with the experiments at IETR's QOSC test facility, which is part of the CNRS RF-Net network.

\bibliographystyle{IEEEtran}


\end{document}